\documentclass[12pt]{article}
\usepackage[dvips]{graphicx}
\usepackage{latexsym}
\parskip=\medskipamount

\def \un{\underline}
\newcommand {\cD}{{\cal D}}

\newcommand {\cF}{{\cal F}}
\newcommand {\cG}{{\cal G}}

\newcommand {\cL}{{\cal L}}
\newcommand {\cM}{{\cal M}}
\newcommand {\cN}{{\cal N}}

\newcommand {\cQ}{{\cal Q}}

\newcommand {\cV}{{\cal V}}
\newcommand {\cW}{{\cal W}}

\newcommand{\bG}{{\bf G}}

\newcommand{\bK}{{\bf K}}
\newcommand{\bL}{{\bf L}}

\newcommand{\bR}{{\bf R}}

\def\a{\alpha}
\def \bi{\bibitem}

\def\b{\beta}
\def\c{\chi}
\def\d{\delta}
\def\e{\epsilon}
\def\f{\phi}
\def\g{\gamma}
\def\G{\Gamma}

\def\k{\kappa}
\def\l{\lambda}
\def\m{\mu}
\def\n{\nu}

\def\p{\pi}
\def\q{\theta}
\def\r{\rho}
\def\s{\sigma}
\def\t{\tau}

\def\x{\xi}
\def\z{\zeta}
\def\D{\Delta}
\def\F{\Phi}
\def\J{\Psi}
\def\L{\Lambda}
\def\O{\Omega}

\def\U{\Upsilon}
\def\X{\Xi}

\newcommand{\ad}{{\dot{\alpha}}}                           
\newcommand{\bd}{{\dot{\beta}}}                            
\newcommand{\ve}{\varepsilon}                            

\newcommand{\pa}{\partial}                           
\newcommand{\hf}{\frac12}

\newcommand{\vf}{\varphi}
\newcommand{\sect}[1]{\setcounter{equation}{0}\section{#1}}

\newcommand{\be}{\begin{equation}}
\newcommand{\ee}{\end{equation}}
\newcommand{\bea}{\begin{eqnarray}}
\newcommand{\eea}{\end{eqnarray}}
\newcommand{\non}{\nonumber}
\begin{document}

\begin{titlepage}
\thispagestyle{empty}

\begin{flushright}
hep-th/0410128 \\
October,  2004 
\end{flushright}
\vspace{5mm}

\begin{center}
{\Large \bf
Self-dual  effective action  of $\cN = 4$ SYM
revisited}
\end{center}

\begin{center}
{\large Sergei M. Kuzenko }
\\
\vspace{2mm}

\footnotesize{
{\it School of Physics, The University of Western Australia\\
Crawley, W.A. 6009, Australia}} \\
{\tt  kuzenko@cyllene.uwa.edu.au}
\\
\end{center}
\vspace{5mm}

\begin{abstract}
\baselineskip=14pt
More evidence is  provided for the conjectured 
correspondence between the D3-brane action 
in $AdS_5 \times S^5$
and the low-energy effective action for $\cN=4$ $SU(N)$ 
SYM on its Coulomb branch,
where the gauge group $SU(N)$ is spontaneously 
broken to $SU(N-1) \times U(1)$ and the dynamics 
is described by a single $\cN=2$ vector multiplet
corresponding to the $U(1)$ factor 
of the unbroken group. 
Using an off-shell formulation for $\cN=4$ SYM
in $\cN=2$ harmonic superspace, within 
the background-field quantization scheme we compute
the two-loop quantum correction to a holomorphic 
sector of the effective action, which is a supersymmetric 
completion of interactions of the form 
$\O \left( (F^+)^2  |Y|^{-4} \right) \,
(F^+)^2(F^-)^2  |Y|^{-4}$, with $F^\pm$ the (anti) 
self-dual components of the $U(1)$ gauge field strength, 
and $Y$ the complex scalar belonging to the vector 
multiplet. In the one-loop approximation, $\O$
was shown in hep-th/9911221  to be constant.
It is  demonstrated  in the present paper that 
$\O \propto (F^+)^2  |Y|^{-4} $  at the two-loop order.
The corresponding coefficient proves to agree with 
the $F^6$ coefficient in the D3-brane action,
after implementing the nonlinear field redefinition 
which was sketched in hep-th/9810152 
and which relates the $\cN=2$ vector multiplet 
component fields with those living on the D3-brane. 
In the approximation considered,  
our results are consistent with the conjecture
of hep-th/9810152 that the $\cN=4$ SYM effective 
action  is self-dual under $\cN=2$ superfield Legendre 
transformation, and also with the stronger 
conjecture of hep-th/0001068 that it is self-dual  
under supersymmetric $U(1)$ duality rotations.  
\end{abstract}

\vfill
\end{titlepage}

\newpage
\setcounter{page}{1}

\renewcommand{\thefootnote}{\arabic{footnote}}
\setcounter{footnote}{0}
\sect{Introduction}
The $\cN=4$ super Yang-Mills (SYM) theory is believed
to be self-dual \cite{MO}.
This was originally formulated as a duality
between the conventional and solitonic
sectors of the theory. More recently, 
inspired in part by the Seiberg-Witten theory 
\cite{SW} and also by the AdS/CFT correspondence 
\cite{Maldacena}, it was suggested  \cite{GKPR} 
that self-duality might be realized in terms of a low-energy 
effective action of the theory on its Coulomb branch, 
where the gauge group $SU(N)$ is spontaneously 
broken to $SU(N-1) \times U(1)$ and the dynamics 
is described by a single $\cN=2$ vector multiplet
corresponding to the $U(1)$ factor 
of the unbroken group. 
Two different implementations of the requirement 
of self-duality for the $\cN=4$ SYM effective action
in $\cN=2$ superspace 
were  proposed :
(i) self-duality under Legendre transformation
\cite{GKPR}; (ii) self-duality under 
$U(1)$ duality rotations \cite{KT}.
So far, neither (i) nor (ii) have been derived 
from first principles, and these proposals 
remain just conjectures. 
Of course, some form  of self-duality 
of the  $\cN=4$ SYM effective action  
is plausible in the context of the AdS/CFT 
correspondence \cite{Maldacena}, 
and this will be discussed below.
Some  concrete support for the
above conjectures  is known to exist 
at the one-loop order.
It turns out that further  support emerges at 
two loops, and this is the main result
of the present paper.   
If the effective action is indeed self-dual 
in the large $N$ limit,  either  
in the sense (i) or (ii),  there should exist  infinitely 
many non-renormalization theorems, and 
this is extremely interesting from the point of view 
of supersymmetric quantum field theory.
Let us therefore start by
recalling   the  self-duality 
equations put forward in \cite{GKPR} and \cite{KT}.

In conventional $\cN=2$ superspace $\bR^{4|8}$ 
parametrized by coordinates 
$z^A = (x^a, \q^\a_i, {\bar \q}_\ad^i)$, 
with $i =\hat{1},  \hat{2}$, 
we denote by $\G[W,\bar W ]$  the effective action of 
$\cN=4$ SYM on its Coulomb branch.
Here the $U(1) $ vector multiplet strength 
$W(z) $ is a chiral superfield, ${\bar D}^i_\ad W =0$, 
satisfying the Bianchi identity \cite{GSW}
\be
D^{ij} W = {\bar D}^{ij} {\bar W}~,
\qquad 
\qquad D^{ij} = D^{\a i} D_\a^{j}~, 
\quad 
 {\bar D}^{ij} 
= {\bar D}_\ad^{i} {\bar D}^{j \ad}~,
\label{ij}
\ee
with $D_A = (\pa_a, D^i_\a, {\bar D}^\ad_i )$
the flat covariant derivatives. 
We  also assume that $\G[W,\bar W ]$
can be unambiguously defined as a functional 
of an unconstrained chiral superfield 
$W$ and its conjugate $\bar W$, and let
\be 
{ {\rm i} \over 2}\, M \equiv  {\d \over \d W} \G[W,\bar W ]~, 
\qquad 
-{{\rm i} \over 2} \, {\bar M} \equiv  {\d \over \d {\bar W}} 
\G[W,\bar W ]
\ee
be the corresponding first variational derivatives. 
${}$Following \cite{Hen}, 
the Legendre transform of $\G[W,\bar W ]$
 is defined by 
\be 
\G_{\rm D}[W_{\rm D} ,{\bar W}_{\rm D} ]
= \G[W,\bar W ]
- {\rm Re}\,  \Big\{ {\rm i} \int {\rm d}^{8}z \, 
W \, W_{\rm D} \Big\}~.
\label{Legendre}
\ee
Here $W_{\rm D}$ is the dual field strength,
\be
W_{\rm D} = {\bar D}^4 D^{ij} \, U_{ij}~, 
\qquad U_{ij}=U_{(ij)}~,
\ee 
with  Mezincescu's  prepotential 
$U_{ij}$   real but otherwise 
unconstrained  \cite{Mez},
and the integration is carried out 
over the chiral subspace  of $\cN=2$ 
superspace.\footnote{Various $\cN=2$ 
superspace integration measures are 
defined in Appendix A.} 
The superfields $W, ~\bar W$ 
on the right of (\ref{Legendre})
have to be expressed via 
$W_{\rm D},~{\bar W}_{\rm D}$ 
using the equation
$M =W_{\rm D} $ and its conjugate.
The requirement of self-duality under Legendre 
transformation introduced in \cite{GKPR} is
\be
\G_{\rm D}[W ,{\bar W} ] =  \G[W ,{\bar W} ] ~.
\label{sde1}
\ee
On the other hand, 
the requirement of invariance  under 
supersymmetric $U(1)$ duality rotations introduced 
in  \cite{KT} is equivalent to the following 
functional equation
\be
{\rm Im}\,  \int {\rm d}^{8}z \, 
\Big\{  W^2 + M^2  \Big\} =0~
\label{sde2}
\ee
which is a generalization of the 
self-duality equation  in nonlinear 
electrodynamics \cite{GR,GZ}.
It can be shown \cite{KT,KT2} 
that the self-duality equation 
(\ref{sde2}) implies (\ref{sde1})
but not vice versa.\footnote{In the case
of nonlinear  electrodynamics, 
the relationship between self-duality 
under Legendre transformation and 
self-duality under $U(1)$ duality rotations 
is analyzed in detail in \cite{IZ}.}

What are physical implications of 
the equations (\ref{sde1}) and (\ref{sde2})?
To answer this question in part, let us
recall the general structure of the low-energy effective 
action for  $\cN=2$ superconformal 
field theories  \cite{BKT}
\be
g^2\, \G = S_{\rm cl}
+ \int {\rm d}^{12}z 
\left\{ c  \ln W \ln {\bar W} 
+ {\rm Re}\, \Big( \L ({\bar \J}) \, \ln W  \Big)
+ \U(\J, {\bar \J}) \right\} 
~+~\cdots
\label{leea}
\ee
where the dots denote higher-derivative terms
as well as those terms which are required by 
quantum modifications of the superconformal 
symmetry.\footnote{The four terms given on the
right of (\ref{leea}) are 
$\cN=2$ superconformal \cite{BKT}.
The complete effective action of $\cN=4$ SYM is 
actually invariant under quantum-corrected 
superconformal transformations \cite{KMT} which differ
from the ordinary linear superconformal  transformations
that leave  the classical action invariant.}
Here 
\be 
S_{\rm cl} =
{\rm Re}\, \Big( \hf 
\int {\rm d}^8z \, 
W^2 \Big)
\label{classical}
\ee
is the classical action for the massless
$U(1)$  vector multiplet,
$\L$ and $\U$ are  holomorphic
and real analytic functions, respectively,
and the (anti)chiral superfields
\be 
{\bar \J} = {\bar W}^{-2} \,D^4 \ln W~, 
\qquad 
 \J = W^{-2} \,{\bar D}^4 \ln {\bar W}
\ee
are  scalar with respect to the 
$\cN=2$ superconformal group 
(see \cite{BKT,KT3} for more details).
Now, the self-duality equation\footnote{If 
the Yang-Mills coupling constant $g\neq 1$,
then $\G$ in eqs.  (\ref{sde1}) and 
(\ref{sde2}) should be replaced by 
$g^2 \G$. Self-duality is compatible with 
$g \neq 1$; if $\G_0[W ,{\bar W} ]$ is a solution 
to (\ref{sde2}), then
$\G [W ,{\bar W} ] =g^2 
\G_0 [W / g , {\bar W}/g ]$ is another solution.}
(\ref{sde2}) implies \cite{KT}, 
in particular,  the following:
\be 
\L ({\bar \J}) = c^2 \, {\bar \J} 
-  c^3 \, {\bar \J}^2 + O ({\bar \J}^3) ~.
\label{f6-f8}
\ee
It is worth pointing out that 
the equation (\ref{sde1}), which 
is weaker than  (\ref{sde2}), 
fixes uniquely the same 
$O({\bar \J} )$ term but leaves 
undetermined the  coefficient for 
the ${\bar \J}^2 $ term \cite{GKPR}.
As regards the  function
$\U(\J, {\bar \J})$ in (\ref{leea}), 
it is not determined completely 
by the self-duality equation 
(\ref{sde2}), since  a general solution  
for this equation
involves a function of one real argument,
 $f(\J{\bar \J})$,  see also \cite{KT2}.
Moreover, the structure of $\U(\J, {\bar \J})$
crucially depends on those terms which are
indicated by the dots in (\ref{leea}). 

The $c$-term\footnote{This functional  
was originally introduced in \cite{dWGR}. 
It is a unique $\cN=2$ superconformal 
invariant  in the family of non-holomorphic actions 
of the form $\int {\rm d}^{12}z \,H(W, {\bar W})$
introduced for the first time in \cite{Hen}.}
 in (\ref{leea}) is known to generate
four-derivative quantum corrections
at the component level; 
these include  an $F^4$ term, 
with $F$ the $U(1)$ field strength.
It is  believed to be generated only 
at one loop in the $\cN=4$ SYM theory 
\cite{DS} (see also \cite{KM5,KM7}), 
in particular it is known that
non-perturbative 
$F^4$ quantum corrections
do not occur in this theory\footnote{Not much 
is known about the structure of instanton 
corrections to the $F^6$ and higher order terms
in $\cN=4$ SYM. But we are interested here  in 
the large $N$ limit in which the instanton 
corrections are subleading.}
\cite{Matone,DKMSW}. 
The explicit value of
$c$  was computed 
by several groups
\cite{GKPR,PvU,G-rR,BK,BBK,LvU}
\be
c = \frac{ (N-1) g^2 }{(4\p)^2} ~.
\label{c}
\ee 
Then, it follows from here and 
eq. (\ref{f6-f8}) that  the $O({\bar \J} )$ term in 
the Taylor expansion of $\L({\bar \J})  $ 
(this term  produces  six-derivative quantum corrections
including $F^6$)
should be generated at two loops only, 
while the ${\bar \J}^2 $ term in 
$\L({\bar \J})  $ (which generates 
special eight-derivative quantum corrections)
should be three-loop exact.

These conclusions  follow 
essentially from 
the self-duality equation, since  
the relation (\ref{c})  is the only input made 
regarding  the quantum $\cN=4$ SYM theory. 
It is natural to wonder to what extent 
these conclusions are supported 
by the quantum theory.  One-loop 
calculations \cite{BKT,KM1,KM7} 
give\footnote{The absence of one-loop
$F^6$ quantum corrections in $\cN=4$ SYM 
was demonstrated years ago in \cite{FT}.}
\be 
\L_{\rm one-loop}({\bar \J}) =0~.
\label{l-one}
\ee
It will be shown below that 
\be 
\L_{\rm two-loop}({\bar \J}) =c^2 \, {\bar \J}~,
\label{l-two}
\ee
with $c$ given in eq. (\ref{c}).
The results (\ref{l-one}) and  (\ref{l-two})
are clearly consistent with  (\ref{f6-f8}). 

The main source of inspiration for the
belief that self-duality of the $\cN=4$ 
SYM effective action is natural, comes 
from the AdS/CFT correspondence.
The latter predicts 
\cite{Maldacena,CT,Tseytlin,BPT}
(a more complete list  of references 
can be found in \cite{BPT}) that the $\cN=4$ 
SYM effective action is related to the 
D3-brane action in $AdS_5 \times S^5$
\bea
S&=& T_3 \int {\rm d}^4x \left( 
h^{-1} - \sqrt{ -\det ( g_{mn}  
+ F_{mn} ) } \right)~, \non \\
g_{mn} &=& h^{-1/2} \eta_{mn}
+h^{1/2}\,  \pa_m X^I \pa_n X^I ~,
\qquad h = {Q  \over (X^I X^I)^2 }~,
\label{D3}
\eea
where $X^I$, $I=1,\cdots, 6$, are transverse 
coordinates,  
$T_3 =(2\p g_s)^{-1}$ 
and  $Q= g_s (N-1)/ \p$.
The action $S/T_3$ is self-dual in the sense 
that it enjoys invariance under 
electromagnetic $U(1)$ duality rotations 
\cite{GR,GZ} (as a consequence of this invariance, 
the action  is also automatically self-dual under Legendre 
transformation \cite{GZ}).
This self-duality of the D3-brane action 
is a fundamental property 
related to the S-duality of type IIB string theory
\cite{TGG}.
If the $\cN=4$ SYM effective action and 
the D3-brane action in $AdS_5 \times S^5$
are indeed related, the former should possess 
some form of self-duality.

With the standard identification $g^2 = 2\p g_s$,
what kind of relationship (in the large $N$ limit) 
should be expected between the bosonic sector 
of (\ref{leea}) and the action (\ref{D3})?
If one  switches off the auxiliary and fermionic 
fields and reduces the action  (\ref{leea})
to  components, should it then coincide with 
(\ref{D3}) assuming that we keep in (\ref{D3})
only two transverse coordinates
$X$'s? The answer is ``No'' 
if we are guided by considerations of  
self-duality \cite{GKPR}.  
Indeed, the $U(1)$ duality rotations 
(or the relevant Legendre transformation)
do not act on the transverse variables 
$X^I$ in (\ref{D3}).  On the other hand, 
the physical  scalar fields $Y = W|_{\q =0}$ 
and $\bar Y$, which belong to the $\cN=2$ vector 
multiplet, do transform under supersymmetric 
$U(1)$ duality rotations \cite{KT} 
(or under the corresponding superfield Legendre 
transformation described above).
The fields $X^I$ and $Y, ~{\bar Y} $ turn out to be 
related to each other by a nonlinear change of 
variables,\footnote{The gauge fields 
corresponding to the actions 
(\ref{leea}) and (\ref{D3})
are also related by a nonlinear change of variables
involving the scalar fields \cite{GKPR}.
The field redefinition under consideration 
was actually worked out 
in \cite{GKPR} in terms of $\cN=1$ superfields.}  
\be
X^I = f^I (\vf , \pa_m \vf, \pa_m \pa_n \vf, \dots)~, 
\quad  I=1,2~, \qquad \quad 
\vf = (Y, {\bar Y } , F_{mn})~,  
\label{change}
\ee
which involves the gauge field strength, 
see also eq. (\ref{comptr}).
This transformation was worked out 
to some order in the derivative expansion
in \cite{GKPR} from completely different, 
duality-independent, 
considerations -- the elimination of 
higher derivatives at the component level. 
The latter point will be discussed a bit later.
But first of all we would like to touch upon one 
particular outcome of the calculation carried out
in \cite{GKPR}. 

In the low-energy  effective action (\ref{leea}), 
let us pick the $\cN=2$ supersymmetric $F^6$ term 
\be 
{\k \over2} \int {\rm d}^{12}z \,
 { \ln W \,
D^4 \ln W \over {\bar W}^2} ~+~{\rm c.c.}~
\label{F6}
\ee
With $c$ given in eq. (\ref{c}),
there are two ``natural'' values for $\k$:
\be  
\k_1 = c^2~;
\qquad \qquad 
\k_2 = {1\over 3} \,c^2~.
\label{kappa}
\ee
The choice $\k=\k_1$ is dictated  
by both self-duality equations 
(\ref{sde1}) and (\ref{sde2}).
On the other hand, 
the choice $\k=\k_2$ corresponds to  
another natural  condition \cite{BPT}, 
namely that, at the component level,  
the effective action (\ref{leea}) should 
produce exactly the same $F^6$ term
as the one contained in the D3-brane action  
(\ref{D3}), if one simply identifies $X^I$ with 
$Y$ and its conjugate 
(setting, in particular,  $X^I X^I = |Y|^2$) 
without any nonlinear redefinition (\ref{change})
implemented. As  is clear, in the self-dual case, 
$\k=\k_1$, one does not generate a correctly 
normalized $F^6$ term in the effective action 
if the dynamical variables are identified with 
the original component fields 
$\vf = (Y, {\bar Y } , F_{mn})$
(appearing in the $\q$-expansion) of the vector multiplet. 
However, the change of variables (\ref{change}) 
constructed in \cite{GKPR}
is such that its implementation in the $c$-term 
generates an additional $F^6$ contribution. 
It turns out that the combined  $F^6$ coefficient 
matches exactly the one in the D3-brane action!

The choice $\k =c^2/3$ is obviously incompatible 
with ($\cN=2$ supersymmetric) self-duality.
It was argued in \cite{BPT} that 
the two-loop $F^6$ quantum correction in 
$\cN=4$ SYM is generated precisely with this 
coefficient. Were this conclusion  correct, 
it would ruin the concept of self-duality 
of the $\cN=4$ SYM effective action.
An independent calculation of the two-loop 
$F^6$ quantum correction in $\cN=4$ SYM 
will be given in the present paper. 
It will lead to the value $\k = c^2$
that respects self-duality.

Apart from considerations of self-duality, 
there is a simple reason as to why one
cannot naively identify the  fields in the 
$\cN=4$ SYM effective action with those 
living on the D3-brane \cite{GKPR}. 
Let us restrict ourselves to the  
consideration of  the 
two- and four-derivative part of  
the effective action 
\be
g^2\, \G =  {\rm Re}\, \Big( \hf 
\int {\rm d}^8z \,  W^2 \Big)
+ c\int {\rm d}^{12}z \,  \ln W \ln {\bar W} 
+\dots
\label{leading}
\ee
At the component level, the $c$-term  
is known to generate not only the structures 
present in the D3-brane action (\ref{D3}), but also 
some terms with higher derivatives. 
There exists no manifestly $\cN=2$ supersymmetric 
functional that could be used to eliminate 
the higher-derivative terms while 
keeping intact the terms present in (\ref{D3}).
The only option remaining to relate the 
effective action to the D3-brane action is 
to resort to a field redefinition of the form (\ref{change}).
This was the approach pursued in \cite{GKPR}.

In relation to the nonlinear field redefinition advocated
in \cite{GKPR}, it is also worth pointing out the following.
${}$For all the known  manifestly supersymmetric 
Born-Infeld (SBI) actions (that is,  
models for partially broken  supersymmetry 
with a vector Goldstone multiplet), 
there is no way to avoid 
a nonlinear field redefinition at the component level 
if one desires to bring the component action to 
a form free of higher derivatives.
Consider, for instance, the $\cN=1$  
SBI action \cite{CF,BG,RT}. 
When reduced to components, 
it automatically leads to the Born-Infeld action in 
the bosonic sector, but the fermionic action turns out 
to contain higher derivatives and, 
as a result, does not coincide 
with the Akulov-Volkov action \cite{VA}.
However, there exists a nonlinear field
redefinition \cite{HK,KMcC} that brings the fermionic action 
to the Akulov-Volkov form. These spin-1 and spin-1/2
features show up again  in the case of 
the $\cN=2$ SBI action\footnote{References to the 
earlier proposals for   $\cN=2$ SBI action
can be found in \cite{KT2}.}
\cite{KT2,BIK}.
In addition,  a new phenomenon occurs 
in the spin-0 sector 
that was not present in the $\cN=1$ case.
The scalar action, which is derived from the
$\cN=2$ SBI theory, contains higher derivatives, 
and a nonlinear field redefinition is required to bring 
the complete bosonic action  to the Dirac-Born-Infeld
form.  The latter follows actually 
from a very simple observation. 
As explained in \cite{KT2}, the $\cN=2$ SBI action 
is a unique solution of the self-duality equation 
(\ref{sde2}) possessing a nonlinearly realized central
charge symmetry. Since the central charge transformation 
is nonlinear, the bosonic action cannot coincide 
with the Dirac-Born-Infeld  action  
\be
S= \int {\rm d}^4x \left( 
1 - \sqrt{ -\det ( \eta_{mn}  
+ \pa_{m} X^I \pa_{n} X^I +
F_{mn} ) } \right)~, 
\label{DBI}
\ee
for the latter  is invariant under linear shifts 
$X^I \to  X^I+ \s^I$, with $\s^I$ constant.

In accordance with our discussion, 
the necessity of  nonlinear field redefinition 
at the component level, is what is common for both 
the $\cN=4$ SYM effective action and 
the SBI actions. There is, however, a fine 
difference\footnote{There also exists an important 
difference between the D3-brane actions in 
$AdS_5 \times S^5$ and Minkowski space. 
In the case of  (\ref{DBI}), any 
field configuration of the form 
$F_{mn}={\rm const}$ and $X^I ={\rm const} $
is a solution of the equations of motion.
In the case of  (\ref{D3}), on the other hand, 
the only constant field 
solutions are $F_{mn}=0$ and $X^I ={\rm const} $.} 
that is due to the fact that the complex scalar  
field in $\cN=4$ has a non-vanishing v.e.v.
In the case of the $\cN=1,\,2$ SBI theories, 
the component action simply coincides with the
Born-Infeld action, if one only keeps the $U(1)$ 
gauge field and switches off the other component 
fields. In the case of the $\cN=4$ SYM effective action, 
however, one does not generate a Born-Infeld 
Lagrangian if the only non-vanishing components 
are the $U(1)$ gauge field and {\it constant} 
scalars. But the Born-Infeld form should be generated
after the field redefinition (\ref{change}).
To a leading order,
the redefinition of \cite{GKPR} 
looks schematically like (with $X=X^1 + {\rm i} \,X^2$)
\be 
X = Y \left( 1  +c 
\, \frac{(F^+)^2}{\g \,|Y|^4} \right) ~+~ 
\dots~,
\label{comptr}
\ee
with $c$ given  in  (\ref{c}),  
$ \g $ a uniquely determined 
numerical coefficient, 
and $F^\pm$ the (anti) self-dual components 
of the $U(1)$ gauge field strength.
Implementing  this in the $F^4$ term 
$$ 
\frac{(F^+)^2 (F^-)^2}{|Y|^4}
$$ 
produces an $F^6$ contribution, in addition to 
the $F^4$ term.

The above consideration relates 
the $\cN=4$ SYM effective action 
(\ref{leea}) to the bosonic D3-brane action 
(\ref{D3}). What about a manifestly supersymmetric 
extension of (\ref{D3}) with two transverse variables
$X^I$ kept? 
As advocated in \cite{KM9} (see also \cite{BIK2}),
it should be a Goldstone multiplet action 
for partially broken 4D $\cN=2$ 
superconformal symmetry associated with 
the coset space $SU(2,2|2) /(SO(4,1) \times 
SU(2))$  and with the dynamics  described
in terms of  a single $\cN=2$ vector multiplet. 
The bosonic sector  of such a model is expected  
to coincide with the D3-brane action in 
$AdS_5 \times S^1$. The structure of 
the corresponding nonlinearly 
realized superconformal  transformations 
is believed  to be related to that of the quantum corrected 
superconformal symmetry in the $\cN=4$ SYM 
theory on its Coulomb branch \cite{KMT}.

Let us now turn to the technical part of 
the present paper.
The holomorphic part of the action
(\ref{leea}) admits a nice representation 
in harmonic superspace  (defined in more detail  
in the next section)
\bea
 \int {\rm d}^{12}z \,  \L ({\bar \J}) \, \ln W 
&=&  \int {\rm d}^{12}z  \int {\rm d}u_1  {\rm d}u_2 \,
\frac{L^{++}(z,u_1) \, 
L^{++}(z,u_2)}
{(u^+_1u^+_2)^2}~,  \non \\
L^{++}(z,u) &=&-  {1 \over 4} (D^+)^2
\Big\{ {\bar W}^{-1} \ln W \,
\sqrt{ \L ({\bar \J}) / {\bar \J}  } \Big\} ~.
\label{holomor}
\eea
The expression for $L^{++}$ can be rewritten 
in the form 
\be
L^{++}(z,u) = {1 \over 4} \,
\frac{(D^+ W)^2}{ W^2 {\bar W}} \,
\sqrt{ \L ({\bar \J}) / {\bar \J}  } 
 ~+~ O\Big( (D^+)^2 W \Big)
\ee
which is quite useful for actual calculations.

This paper is organized as follows.
Section 2 contains the necessary setup
regarding the $\cN=4$ super Yang-Mills  
theory  and its background-field quantization 
in $\cN=2$ harmonic superspace. 
In section 3 we describe the exact heat kernels
in $\cN=2$ superspace, which correspond 
to various covariantly constant vector multiplet 
backgrounds.  Section 4, the central part of this work,
is devoted to the evaluation 
of the two-loop quantum corrections to $\L({\bar \J})$ 
in (\ref{leea}). Discussion and conclusions 
are given in section 5. The main body of the paper 
is accompanied by four appendices. 
In appendix A we define all the  $\cN=2$ superspace 
integration measures used throughout this paper. 
Appendix B describes the Cartan-Weyl 
basis for  $SU(N)$ used in the paper.  
In appendix C  the main properties
of the parallel displacement propagator 
are given. An alternative representation for the free
massless propagators in harmonic superspace is given 
in appendix D.

\sect{\mbox{$\cN = 4$} SYM setup}

Throughout this paper, the $\cN=4$ SYM theory is 
treated as $\cN=2$ SYM coupled to a hypermultiplet 
in the adjoint representation of the gauge group.  
We therefore start by assembling 
the necessary information about  
the $\cN=2$ Yang-Mills supermultiplet
which is known to possess an off-shell formulation 
\cite{GSW} in 
conventional $\cN=2$ superspace $\bR^{4|8}$, 
which turns out to be a gauge fixed version 
of an off-shell  formulation \cite{GIKOS}
in $\cN=2$ harmonic superspace 
$\bR^{4|8} \times S^2$.

\subsection{The $\cN=2$ super Yang-Mills geometry}
In order to describe 
the $\cN=2$ Yang-Mills  supermultiplet, 
one introduces, following \cite{GSW}, 
superspace  gauge covariant derivatives defined by  
\be 
\cD_A = (\cD_a, \cD^i_\a, {\bar \cD}^\ad_i )
= D_A + {\rm i}\,  \cV_A (z)~,
\label{gcd}
\ee 
with $D_A = (\pa_a, D^i_\a, {\bar D}^\ad_i )$
the flat covariant derivatives, and $\cV_A$
the gauge connection.
Their gauge transformation law is 
\be
\cD_A ~\to ~ {\rm e}^{{\rm i} \t(z)} \, \cD_A\,
{\rm e}^{-{\rm i} \t(z)}~, \qquad 
\t^\dagger = \t ~, 
\label{tau}
\ee
with the gauge parameter $\t(z)$ being arbitrary modulo 
the reality condition imposed.
The gauge covariant derivatives
obey the following algebra \cite{GSW}:
\bea 
\{\cD^i_\a , {\bar \cD}_{\ad j} \}
&=& - 2{\rm i}\,\d^i_j \cD_{\a \ad} \non \\
\{\cD^i_\a , \cD^j_\b \} = 2{\rm i}\, \ve_{\a\b}
\ve^{ij} {\bar \cW}~, \quad && \quad 
\{ {\bar \cD}_{\ad i} , {\bar \cD}_{\bd j} \} 
= 2{\rm i}\, \ve_{\ad \bd} \ve_{ij} \cW~, \non \\
\left[ \cD^i_\a , \cD_{\b \bd} \right]  
= \ve_{\a\b} {\bar \cD}^i_\bd {\bar \cW}~,
 \quad && \quad
[ {\bar \cD}_{\ad i}, \cD_{\b \bd}] 
=\ve_{\ad \bd}\cD_{\b i} \cW ~, \non \\
\left[ \cD_{\a \ad} , \cD_{\b \bd} \right]
= {\rm i}\, \cF_{\a \ad,\b \bd} &=& 
{ {\rm i}\, \over 4} \,  \ve_{\ad \bd}\, 
\cD^i{}_{(\a} \cD_{\b )i} \cW 
- { {\rm i} \over 4}\, \ve_{\a \b}\, 
{\bar \cD}^i{}_{(\ad} {\bar \cD}_{\bd )i} {\bar \cW} ~.
\label{algebra}
\eea
The superfield strengths $\cW$ and $\bar \cW$ satisfy 
the Bianchi identities 
\be
{\bar \cD}^i_\ad \cW = \cD^i_\a {\bar \cW} =0~,
\qquad 
\cD^{ij} \cW = {\bar \cD}^{ij} {\bar \cW}~,
\ee
where 
\be
\cD^{ij} = \cD^{\a (i} \cD_\a^{j)} ~, \qquad 
 {\bar \cD}^{ij} 
= {\bar \cD}_\ad^{(i} {\bar \cD}^{j) \ad}~.
\ee

The $\cN=2$ harmonic superspace 
$\bR^{4|8}\times S^2$
\cite{GIKOS,GIOS1,GIOS2}
extends conventional superspace
by the two-sphere $S^2 = SU(2)/U(1)$
parametrized by harmonics, i.e., group
elements
\bea
({u_i}^-\,,\,{u_i}^+) \in SU(2)~, \quad
u^+_i = \ve_{ij}u^{+j}~, \quad \overline{u^{+i}} = u^-_i~,
\quad u^{+i}u_i^- = 1 ~.
\eea
In harmonic superspace,
both the $\cN=2$ Yang-Mills supermultiplet and
hypermultiplets can be described 
by {\it unconstrained} 
superfields over the analytic
subspace of $\bR^{4|8}\times S^2$
parametrized by the variables
$ \z \equiv (y^a,\q^{+\a},{\bar\q}^+_{\dot\a}, \,
u^+_i,u^-_j) $, 
where the so-called analytic basis\footnote{The
original parametrization of harmonic superspace, 
in terms of the variables  $Z=(z^A, u^+_i,u^-_j)$, 
is known as the central basis.}
 is defined by
\be
y^a = x^a - 2{\rm i}\, \q^{(i}\s^a {\bar \q}^{j)}u^+_i u^-_j~, 
\qquad
 \q^\pm_\a=u^\pm_i \q^i_\a~, \qquad {\bar \q}^\pm_{\dot\a}
=u^\pm_i{\bar \q}^i_{\dot\a}~.
\ee
With the notation 
\be 
 \cD^\pm_\a=u^\pm_i \cD^i_\a~, 
\qquad {\bar \cD}^\pm_{\dot\a}
=u^\pm_i{\bar \cD}^i_{\dot\a}~,
\ee
it follows from (\ref{algebra}) that 
the operators $\cD^+_\a$ and 
${\bar \cD}^+_{\dot\a}$ strictly anticommute, 
\be
 \{ \cD^+_\a , \cD^+_\b \} =
\{ {\bar \cD}^+_{\dot\a} , {\bar \cD}^+_{\dot\b} \}
= \{ \cD^+_\a , {\bar \cD}^+_{\dot\a} \} =0~.
\label{analintegrability}
\ee
A covariantly analytic superfield
$\F^{(p)}(z,u)$ is defined to 
be annihilated by these operators,
\be
\cD^+_\a \F^{(p)} = {\bar \cD}^+_\ad \F^{(p)}=0~. 
\label{analytic-con}
\ee
Here the superscript $p$ refers to the harmonic 
U(1) charge, $\cD^0 \, \F^{(p)} = p\,  \F^{(p)}$, 
where $\cD^0$ is one of the harmonic gauge 
covariant  derivatives
which in the central basis are:
\be 
\cD^0 = u^{+i} \, \frac{\pa} {\pa u^{+i}} 
- u^{-i} \, \frac{\pa} {\pa u^{-i}} 
\equiv D^0 ~,
\qquad
\cD^{\pm \pm} = u^{\pm i} \, 
\frac{\pa} {\pa u^{\mp i}} 
\equiv D^{\pm \pm} 
 ~.
\ee
The operator $\cD^{++}$ acts on the space 
of covariantly analytic superfields.

It follows from (\ref{analintegrability}) that 
\be 
\cD^+_\a = {\rm e}^{-{\rm i}\O} \, D^+_\a \,
{\rm e}^{{\rm i}\O}~, 
\qquad 
{\bar \cD}^+_\ad = {\rm e}^{-{\rm i}\O} \, 
{\bar D}^+_\ad \,
{\rm e}^{{\rm i}\O}~
\ee
for some Lie-algebra-valued superfield 
$\O(z,u) $ known as  the bridge\footnote{The
bridge can be chosen to real with  respect 
to a uniquely defined  
analyticity-preserving conjugation,
$\breve{\O} =\O$, see  
\cite{GIKOS,GIOS2} for more details.} 
\cite{GIKOS,GIOS2}. 
The latter relations inspire one to introduce
the so-called $\l$-frame  defined by 
\be
\cD_A ~ \longrightarrow ~ 
{\rm e}^{{\rm i}\O} \, \cD_A \,
{\rm e}^{-{\rm i}\O}~, \quad 
\cD^{\pm \pm} 
~ \longrightarrow ~ 
{\rm e}^{{\rm i}\O} \, \cD^{\pm \pm} \,
{\rm e}^{-{\rm i}\O}~,
\quad  \F^{(p)} ~ \longrightarrow ~ 
{\rm e}^{{\rm i}\O} \,   \F^{(p)} ~.
\ee 
In the $\l$-frame, the gauge covariant 
derivatives $\cD^+_\a$ and ${\bar \cD}^+_\ad$ 
coincide with the flat derivatives 
$D^+_\a$ and ${\bar D}^+_\ad$, respectively, 
and therefore any  covariantly analytic 
superfield becomes a function over 
the analytic subspace, $\F^{(p)} =\F^{(p)} (\z)$.
Unlike $\cD^+_\a$ and ${\bar \cD}^+_\ad$, 
the gauge covariant derivatives
$\cD^{++}$ and $\cD^{--}$ acquire 
connections, 
\be
\cD^{\pm \pm} = D^{\pm \pm} +
{\rm i}\,\cV^{\pm\pm}(Z)~.
\ee
It can be shown that $\cV^{++}$ is 
an {\it unconstrained analytic} superfield
\cite{GIKOS},  $D^+_\a \cV^{++}=
{\bar D}^+_\ad \cV^{++}=0$,
and the other geometric objects
$\cV_A$ and $\cV^{--}$ are determined in terms
of $\cV^{++}$ \cite{Z}. 
Therefore $\cV^{++}$ is the only independent 
prepotential containing all the information 
about the $\cN=2$ vector  multiplet.
In the $\l$-frame, the gauge transformation law 
(\ref{tau}) turns into 
\be
\cD_A ~\to ~ {\rm e}^{{\rm i} \l(\z)} \, \cD_A\,
{\rm e}^{-{\rm i} \l(\z)}~, \quad 
\cD^{\pm\pm} ~\to ~ {\rm e}^{{\rm i} \l(\z)} \, 
\cD^{\pm\pm}\,
{\rm e}^{-{\rm i} \l(\z)}~, \qquad 
\breve{ \l } = \l ~, 
\label{lam}
\ee
with the analytic gauge parameter $\l$ 
being real, with respect to the analyticity-preserving 
conjugation, but otherwise arbitrary.
The original representation, in which the gauge 
covariant derivative $\cD_A$ are 
harmonic-independent and 
$\cD^{\pm \pm}$ are connection-free, 
is called the $\t$-frame \cite{GIKOS}.
In what follows, we usually do not  specify 
the frame to be used.

\subsection{The background-field quantization 
of $\cN=4$ SYM}

In order to realize the  $\cN=4$  
super Yang-Mills theory in $\cN=2$ 
harmonic superspace, 
we  choose its  classical action,  
$S = S_{\rm SYM} + S_{\rm hyper}$,
to consist of two parts: 
(i) the pure $\cN=2$ SYM action \cite{GSW,Z}
\bea
S_{{\rm SYM}}
&=& \frac{1}{2g^2} {\rm tr}\int 
{\rm d}^8z\, \cW^2=
\frac{1}{2g^2}{\rm tr} 
\int 
{\rm d}^8{\bar z}\, {\bar \cW}^2 
\label{n=2pure-sym}
 \\
&=& \frac{1}{g^2} {\rm tr}
\int {\rm d}^{12}z
\sum\limits_{n=2}^\infty\frac{(-{\rm i})^n}{n}
\int  
\prod_{a=1}^{n}{\rm d}u_a \,
\frac{\cV^{++}(u_1)
\cV^{++}(u_2)
\dots \cV^{++}(u_n)}
{(u^+_1u^+_2)(u^+_2u^+_3)\dots(u^+_nu^+_1)}
~; \non
\eea
 (ii) the $q$-hypermultiplet action \cite{GIKOS}
in the adjoint representation of the gauge group
\be 
S_{{\rm hyper}} = - \frac{1}{g^2} {\rm tr}
\int {\rm d}\zeta^{(-4)}
\breve{Q}{}^+ \,\cD^{++}\,Q^+ ~.
\label{hypermul}
\ee
It is assumed in  (\ref{n=2pure-sym}) 
and (\ref{hypermul}) that the trace is taken   
in the fundamental representation 
of $SU(N)$, $ {\rm tr} ={\rm tr}_{\rm F} $,
in which   the  generators are normalized such that  
${\rm tr} \,(T^\m\, T^\n) = g^{\m \n}$, with 
$g^{\m\n}$ the Cartan-Killling  metric
(see Appendix B).

By construction, 
the  action $S = S_{\rm SYM} + S_{\rm hyper}$
is manifestly $\cN=2$ supersymmetric. 
It also proves to  be invariant under  two hidden
supersymmetries (parametrized by constant 
spinors $\e^\a_i$ and their conjugates)
of the form \cite{GIOS1,BBK}
\bea
\d \cV^{++} = 
(\e^{ i} \,\q^+ + {\bar \e}^i \,{\bar \q}^{+} ) \,
Q^+_i~, 
\quad
\d Q^+_i = - {1 \over 4}(\cD^+)^2 ( \e_i \,\q^- \, \cW)
- {1 \over 4}({\bar \cD}^+)^2 
({\bar \e}_{ i}\, {\bar \q}^{ -} \, {\bar \cW})~,
\eea
with the notation 
$Q^+_i = (Q^+, \breve{Q}{}^+)$.
 
To quantize the theory, we  use the $\cN=2$ 
background field formulation \cite{BBKO,BKO,BK}
(see also \cite{BOBIK} for a review) 
and split the (unconstrained analytic) dynamical variables 
into background and quantum ones, 
\bea
 \cV^{++} ~ \to ~ \cV^{++} +v^{++} ~, 
\qquad \cQ^+ ~ \to  ~ \cQ^+ +q^+ ~, 
\qquad 
\breve{\cQ}^+ ~ \to ~ \breve{\cQ}^++ \breve{q}^+ ~,
\eea
with lower-case letters used for 
the quantum superfields. 
In this paper, we are not interested in the 
dependence of the effective action on 
the hypermultiplet superfields, 
and therefore we set $\cQ^+ = \breve{\cQ}^+ =0$
in what follows.
We also specify the background $\cN=2$ 
vector multiplet to satisfy the classical equations
of motion  
\be
\frac{\d S_{\rm SYM}}{\d \cV^{++} } 
= 0~, \qquad 
g^2 \, \frac{\d S_{\rm SYM} }{\d \cV^{++} } 
={1 \over 4} \,(\cD^+)^2 \cW
={1 \over 4} \,({\bar \cD}^+)^2 {\bar \cW} ~.
\label{on-shell}
\ee
After the background-quantum splitting, 
the action (\ref{n=2pure-sym}) turns into\footnote{To 
simplify the notation,  we set $g^2 =1$ at the intermediate 
stages of the calculation. The explicit dependence 
on the coupling constant will be restored 
in the final expression for the effective action.}
\bea
S_{{\rm SYM}}[\cV^{++}+v^{++}]
=S_{{\rm SYM}}[\cV^{++}] 
+ \D S_{{\rm SYM}}[v^{++},\cV^{++}] ~,
\eea
where the quantum part
\bea
\D S_{{\rm SYM}}
&=&
{\rm tr}\,\int
{\rm d}^{12}z\sum\limits_{n=2}^\infty {\frac{(-{\rm i})}{n}}^{n}
\int \prod_{a=1}^{n}{\rm d}u_a \,
\frac{v^{++}(u_1)v^{++}(u_2)\dots v^{++}(u_n)}
{(u^+_1u^+_2)(u^+_2u^+_3)\dots(u^+_nu^+_1)}
\eea
is given in the $\t$-frame associated with the
background vector multiplet
\cite{BBKO}.
The hypermultiplet action becomes
\be 
S_{{\rm hyper}} = - 
{\rm tr}
\int {\rm d}\zeta^{(-4)} \,\Big\{ 
\breve{q}{}^+ \, \cD^{++}q^+ 
+{\rm i} 
\,\breve{q}{}^+ \,
[v^{++},q^+] \Big\}~.
\label{hyper}
\ee

In accordance with  \cite{BBKO,BKO},
choosing the background-covariant 
gauge condition $\c^{(4)} = \cD^{++} v^{++}$,
and the gauge-fixing functional 
\be
S_{\rm gf} = - \hf
{\rm tr}\, \int {\rm d}^{12}z \int
{\rm d}u_1{\rm d} u_2\,
\c^{(4)} (z,u_1)\,
\frac{(u^-_1u^-_2)}{(u^+_1u^+_2)^3} \,
\c^{(4)}(z,u_2)~,
\ee
one arrives at the following gauge-fixed  
action for the quantum Yang-Mills superfield
\bea
\D S_{{\rm SYM}} &+& S_{\rm gf}  
= \hf {\rm tr}
\int {\rm d}\zeta^{(-4)} 
v^{++} {\stackrel{\frown}{\Box}}v^{++} \non \\
&+&{\rm tr}\,\int
{\rm d}^{12}z\sum\limits_{n=3}^\infty {\frac{(-{\rm i})}{n}}^{n}
\int \prod_{a=1}^{n}{\rm d}u_a \, 
\frac{v^{++}(u_1)v^{++}(u_2)\dots v^{++}(u_n)}
{(u^+_1u^+_2)(u^+_2u^+_3)\dots(u^+_nu^+_1)}
~,
\label{gluon}
\eea
with  ${\stackrel{\frown}{\Box}}$ 
the analytic d'Alembertian \cite{BBKO},
\bea
{\stackrel{\frown}{\Box}}{}&=&
{\cal D}^a{\cal D}_a -
\frac{{\rm i}}{2}({\cal D}^{+\a}\cW){\cal D}^-_\a
-\frac{{\rm i}}{2}
({\bar{\cal D}}^+_{\dot\alpha}{\bar \cW}){\bar{\cal D}}^{-{\dot\alpha}}
+\frac{{\rm i}}{4}({\cal D}^{+\a} {\cal D}^+_\a \cW) \cD^{--}\non \\
&{}& -\frac{{\rm i}}{8}[{\cal D}^{+\alpha},{\cal D}^-_\alpha] \cW
- \frac{1}{2}\{{\bar \cW},\cW \}~.
\label{analdal}
\eea
${}$Furthermore, the gauge condition chosen leads
to the Faddeev-Popov ghost action 
\be 
S_{\rm ghost} = -
{\rm tr} \, \int {\rm d}\zeta^{(-4)}\, \Big\{
(\cD^{++} b)\,
\cD ^{++}c + 
{\rm i}\, (\cD^{++} b) \,
[v^{++},c]\Big\}~,
\label{ghost}
\ee
where the ghosts $b$ and $c$ 
are background covariantly analytic superfields.

Along with the Faddeev-Popov ghosts,  $b$ and $c$, 
one should also take into account several 
Nielsen-Kallosh ghosts
for the theory under consideration \cite{BBKO,BKO}. 
But the latter turn out to contribute at the one-loop 
order only,\footnote{The correct
structure of the Nielsen-Kallosh ghosts for 
$\cN=2$ SYM theories in harmonic superspace
is given  in \cite{BKO}. 
The one-loop low-energy 
effective action for $\cN=4$ SYM was evaluated 
in \cite{BK,BBK,KM1}.}
and therefore  we do not spell out their 
explicit form in the present paper.

With the condensed notation $Z=(z^A, u^+_i,u^-_j)$, 
the quadratic parts of the actions 
(\ref{hyper}), (\ref{gluon}) and (\ref{ghost}) 
define the Feynman propagators 
\bea 
{\rm i} \,\langle 
{ q}^+_\m(Z)\,\breve{ q}{}^+_{\n '}(Z') \rangle 
&=& G^{(1,1)}_{\m \, \n '} (Z , Z')~, 
\qquad G^{(1,1)}_{\m \, \m'} (Z , Z')
=-G^{(1,1)}_{\m' \,  \m} (Z' , Z)~,
\non \\
{\rm i} \,\langle 
{v}^{++}_\m(Z)\, v^{++}_{\n'}(Z') \rangle 
&=& G^{(2,2)}_{\m \, \n'} (Z , Z')~, 
\qquad G^{(2,2)}_{\m \, \m'} (Z , Z')
=G^{(2,2)}_{\m' \, \m} (Z' , Z)~,
\\
{\rm i} \,\langle 
c_\m(Z)\,b_{\n'}(Z') \rangle 
&=& G^{(0,0)}_{\m \, \n'} (Z , Z')~,
\qquad G^{(0,0)}_{\m \,\m'} (Z , Z')
=G^{(0,0)}_{\m' \, \m} (Z' , Z)
~. \non
\eea
Here the Green functions 
$G^{(0,0)} (Z , Z')$, 
$G^{(1,1)} (Z , Z')$ and $G^{(2,2)} (Z , Z')$
are background covariantly analytic 
in both arguments.
They satisfy the equations 
\bea 
(\cD^{++})^2 \,G^{(0,0)} (Z , Z') &=& 
-\d^{(4,0)}_{\rm A} (Z,Z')~, \non \\
\cD^{++} \,G^{(1,1)} (Z , Z') &=& 
\d^{(3,1)}_{\rm A} (Z,Z')~, \\
{\stackrel{\frown}{\Box}}{}\,G^{(2,2)} (Z , Z')
&=&  - \d^{(2,2)}_{\rm A} (Z,Z')~, \non
\eea
where 
\bea 
\d^{(4-n,n)}_{\rm A} (Z,Z')
&=& (\cD^+)^4 
\Big\{ {\bf 1}\, \delta^{12}(z-z')\,
\d^{(-n,n)} (u,u')\Big\} \non \\
&=&(\cD'^+)^4 
\Big\{ {\bf 1}\, \delta^{12}(z-z')\,
\d^{(4-n,n-4)} (u,u')\Big\}
\eea
is a background covariantly analytic
delta-function.
The explicit form of the Green functions is 
as follows \cite{BBKO}:
\bea
G^{(0,0)} (Z, \,  Z')
&=& - (u^-u'^-)\,G^{(1,1)} (Z,Z')
~, \non \\
G^{(1,1)} (Z,Z')
&=& \phantom{-}
\frac{1}{{\stackrel{\frown}{\Box}}{}}\,
(\cD^+)^4 \,(\cD'^+)^4\,
\Big\{ {\bf 1}\, \delta^{12}(z-z')\,
{1 \over (u^+ u'^+)^3}\Big\}~,  
\label{green}\\
G^{(2,2)} (Z,Z')
&=& -\frac{1}{{\stackrel{\frown}{\Box}}}\,
(\cD^+)^4 
\Big\{ {\bf 1}\, \delta^{12}(z-z')\,
\d^{(-2,2)} (u,u')\Big\}~. \non
\eea
Switching off the background vector multiplet, 
these Green functions reduce to the free
massless ones \cite{GIOS1,GIOS2,OY}.

\subsection{Specification of the background vector multiplet}

We are interested in quantum dynamics 
on the Coulomb branch of $\cN=4$ SYM.
More specifically, the effective action will be computed 
only for a $U(1)$ vector multiplet corresponding 
to a special direction in the Cartan subalgebra 
of $SU(N)$. With the Cartan-Weyl basis for $SU(N)$
defined in Appendix B,  eqs. (\ref{C-W}) and (\ref{C-W-2}),
the  background vector multiplet  
will be chosen 
\be
\cW = W \, H^0~, 
\label{actual-background}
\ee
with $W$ the $U(1)$ gauge field strength
possessing a nonzero v.e.v.    
Its characteristic feature is that it leaves
the subgroup $SU(N-1) \times U(1) \subset SU(N)$ 
unbroken, where $U(1)$ is associated with $H^0$
and  $SU(N-1)$ is  generated by 
$\{ H^{\un{I}}, \, E^{ \un{i} \un{j}} \}$.

The mass-like term in 
the analytic d'Alembertian
(\ref{analdal}) becomes 
\be
\frac{1}{2}\{{\bar \cW},\cW \}=
{\bar \cW} \cW = {\bar W}W \, (H^0)^2~, 
\ee
and therefore a superfield's mass is determined 
by  its $U(1)$ charge with respect to $H^0$. 
With the notation 
\be
e = \sqrt{ N /( N -1)}  ~,
\ee
the $U(1)$ charges of the components
of a quantum superfield $v_\m$
are given in the table.
\begin{center}
\begin{tabular}{ | c || c | c | c | c |}  \hline
superfield &  
$v_{0\, \un{i} }$ & $v_{\un{i} \, 0}$ & 
$v_I$ & $v_{ \un{i} \,\un{j} }$\\ \hline 
$U(1) $ charge & 
$e $ & $-e $ 
&0&0\\  \hline
\end{tabular} \\
${}$\\
Table  1: $U(1)$ charges of superfields
\end{center}
Here $v_\m$ stands for any of 
the quantum superfields $v^{++}_\m, ~
q^+_\m, ~ b_\m $ and $c_\m$. 
Among the components of $v_\m$,
there are $2(N-1)$ charged superfields 
($v_{0 \un{i}}$ and their conjugates $v_{ \un{i}0}$)
coupled to the background, 
while the remaining  $(N-1)^2$ 
neutral superfields ($v_I$ and 
$v_{\un{i} \,\un{j}} $)  
do not interact with  the background 
and, therefore, are free massless.
This follows from the identity 
\be
[H^0 , E^{ij}] ~=~ \sqrt{N \over N-1}\,  \Big( 
\d^{0i}\, E^{0j} - \d^{0j}\, E^{i0} \Big)~. 
\ee

${}$For the background chosen, 
the Green function
$\cG =(\cG_\m{}^\n)$, with 
\be
\cG_\m{}^\n (Z,Z') = G_{\m \l}(Z,Z') \, g^{\l \n} ~, 
\qquad  G_{\m \n}(Z,Z') = {\rm i} 
\langle v_\m (Z) \, v_\n (Z') \rangle~,
\label{G}
\ee
is diagonal. Relative to the basis 
$T^\m = (H^I, E^{0 \un{i} }, E^{\un{i} 0} , E^{ \un{i} \un{j} })$, 
this Green's function has the form
\be
\cG = {\rm diag} \Big(\bG_{\{0\}}\, {\bf 1}_{N-1} , ~
\bG_{\{e \}}\, {\bf 1}_{N-1}, ~
\bG_{\{-e \}}\, {\bf 1}_{N-1}, ~
\bG_{\{0\}}\, {\bf 1}_{(N-1)(N-2)} \Big)~.
\label{I+II-ad12}
\ee
Here $\bG_{\{e\}} (Z,Z') $ denotes a $U(1)$
Green function of charge $e$.

\sect{Heat kernel in  covariantly constant backgrounds}

As demonstrated in \cite{K}, 
in the case of an on-shell 
background vector multiplet, 
eq. (\ref{on-shell}), 
the analytic d'Alembertian in 
the Green functions (\ref{green}) can be 
replaced by the following harmonic-independent 
operator
\bea 
\D &=& {\stackrel{\frown}{\Box} }
 + \frac{{\rm i}}{2}({\cal D}^{-\a}\cW){\cal D}^+_\a
+\frac{{\rm i}}{2}
({\bar{\cal D}}^-_\ad{\bar \cW}){\bar{\cal D}}^{+\ad} \non \\
&=& \cD^a \cD_a 
+ \frac{{\rm i}}{2}
(\cD^\a_i \cW)\cD_\a^i
-\frac{{\rm i}}{2}
({\bar \cD}^i_\ad{\bar \cW}){\bar \cD}^\ad_i
- \frac{1}{2}\{{\bar \cW},\cW \}~.
\label{Delta}
\eea
Then, all information about the Green functions 
is actually encoded in the superfield heat kernel 
\be
K(z,z'|s) = {\rm e}^{{\rm i} s \D} \,
\Big\{ {\bf 1} \, \d^{12}(z-z')\Big\} ~.
\label{heat-kernel}
\ee
Indeed, the Green functions are now defined 
by the Fock-Schwinger proper-time representation
\bea
G^{(1,1)} (Z,Z')
&=&  -{\rm i}  
\int_{0}^{\infty} {\rm d}s \, {\rm e}^{-\ve \, s}\,
(\cD^+)^4 (\cD'^+)^4\, {K(z,z'|s)   
 \over (u^+ u'^+)^3}~,  
 \non \\
G^{(2,2)} (Z,Z')
&=& \phantom{-}
{\rm i}  
\int_{0}^{\infty} {\rm d}s \, {\rm e}^{-\ve \, s}\,
(\cD^+)^4 
K(z,z'|s) \,  \d^{(-2,2)} (u,u') ~,
\label{short}
\eea
with $ \ve \to +0$.
In what follows, the damping factor
$ {\rm e}^{-\ve \, s}$ 
is always assumed in the proper-time 
integrals, but not given explicitly.  

The gluon Green function, 
$G^{(2,2)}$,  can be  
represented\footnote{Use of the $\t$-frame is assumed
in the second line of (\ref{longer}).}
in a manifestly analytic form \cite{K}
(see also \cite{GKS,GIOS2})
\bea
G^{(2,2)} (Z,Z')
&=&  - \hf (\cD^+)^4 (\cD'^+)^4   
(\cD^{--})^2 \, 
\frac{1}{\D^2}\,
\Big\{ {\bf 1}\, \delta^{12}(z-z')  \,
\d^{(2,-2)} (u,u') \Big\}\non \\
&=& (\cD^+)^4 (\cD'^+)^4  \left\{ 
\hf \int_{0}^{\infty} {\rm d}s \, s\,
K(z,z'|s) \,  (\cD^{--})^2 \, \d^{(2,-2)} (u,u') 
\right\}~,
\label{longer}
\eea
which may be useful when computing 
supergraphs involving products of 
harmonic distributions (see below).
This representation is often 
known\footnote{It is only the 
{\it free} superpropagators
which are analyzed in \cite{GIOS1,GIOS2}.}
\cite{GIOS2} as the ``longer form'' of the 
propagator $G^{(2,2)}$, while its original representation, 
eqs. (\ref{green}) and (\ref{short}), 
is called the ``short form'' of  $G^{(2,2)}$. 

The heat kernel can be computed exactly \cite{K}
in the case of a covariantly constant vector multiplet, 
\be 
\cD_a \cW = \cD_a {\bar \cW} =0 
\quad \Longrightarrow \quad 
[\cW, {\bar \cW}] =0~,
\ee
and the result is\footnote{The 
determinant in (\ref{heatkernel1})
is computed with respect to the
Lorentz indices.}
\bea
K(z,z'|s) &=& -\frac{\rm i}{(4 \pi s)^2} \, 
{\bf det} \left( \frac{ s \cF}
{\sinh  ( s \cF) }
\right)^{1 \over 2}
\, {\rm e}^{ \frac{{\rm i}}{4} \r(s)
 \cF \coth (s \cF) \r (s) }
\, \d^4(\x (s) )\, \d^4(\bar{\x}(s)) \,
 {\rm e}^{{\rm i} s \U} I(z,z')~, \non \\
&& 
\qquad \qquad 
\X^A(s) = {\rm e}^{{\rm i} s \U}  \, \X^A \,
{\rm e}^{-{\rm i} s \U} ~, 
\qquad \X^A =(\r^a, \x^\a_i , {\bar \x}_\ad^i)~,
\label{heatkernel1}
\eea 
with 
\be 
\U = \frac{{\rm i}}{2}
(\cD^\a_i \cW)\cD_\a^i
-\frac{{\rm i}}{2}
({\bar \cD}^i_\ad{\bar \cW}){\bar \cD}^\ad_i             
-{\bar \cW} \cW 
\ee
 being the first-order operator that appears in  (\ref{Delta}).
Here we have introduced the $\cN=2$ 
supersymmetric  interval
$\X^A \equiv \X^A (z,z') = -\X^A (z',z) $ 
defined by
\bea
 \X^A = 
\left\{
\begin{array}{l}
\r^a = (x-x')^a - {\rm i} (\q-\q')_i \s^a {\bar \q}'^i 
+ {\rm i} \q'_i \s^a ( {\bar \q} - {\bar \q}')^i ~, \\
\x^\a_i = (\q - \q')^\a_i ~, \\
{\bar \x}_\ad^i =({\bar \q} -{\bar \q}' )_\ad^i ~. 
\end{array} 
\right. 
\label{super-interval}
\eea
The parallel displacement propagator, 
$I(z,z') $, and its properties \cite{KM3} are collected 
in the appendix. It is worth pointing out that
\be
[\U, \cW] = [\U, {\bar \cW}] =0~,
\ee
and therefore
\be
{\rm e}^{{\rm i} s \U} = {\rm e}^{- {\rm i} s {\bar \cW} \cW }\,
{\rm e}^{ {\rm i} s A}~, 
\qquad  A= \frac{{\rm i}}{2}
(\cD^\a_i \cW)\cD_\a^i -\frac{{\rm i}}{2}
({\bar \cD}^i_\ad{\bar \cW}){\bar \cD}^\ad_i     
~.
\ee   
The components of $\X^A(s) $
can be easily evaluated using the (anti) commutation
relations (\ref{algebra}) and   
the obvious identity
\be 
D_B \, \X^A = \d_B{}^A 
+\hf \, \X^C \, T_{CB}{}^A~,
\label{trivial}
\ee
with $T_{CB}{}^A$ the flat superspace torsion.

With the notation 
\be
\cM_\a{}^\b = -{1\over 4} \cD^i_{\a} \, \cD^\b_i \,
\cW~,  \qquad \cM_\a{}^\a =0~,
\ee
one readily computes
\be
\x^\a_i (s) = \x^\a_i -\hf\Big( \cD_i\cW \, 
\frac{ {\rm e} ^{s\cM} -{\bf 1} } {\cM} \Big)^\a~,
\ee
and therefore
\be
\cD^j_\b \,\x^\a_i (s)  = \d^j_i \, 
({\rm e} ^{s\cM})_\b{}^\a~.
\ee
The latter identity is equivalent\footnote{Here and below, 
it  is understood that $\x^\pm_\a = \x^i_\a \,u^\pm_i$, 
$\x'^\pm_\a = -\x^i_\a \,u'^\pm_i$, and similarly 
for $\bar \x$'s.}
to 
\bea
\cD^+_\b \x^{+\a} (s)= \cD^-_\b \x^{-\a} (s) =0~, 
\qquad 
\cD^+_\b \x^{-\a} (s)= - \cD^-_\b \x^{+\a} (s) =
({\rm e} ^{s\cM})_\b{}^\a~,
\label{fun1}
\eea
and hence 
\be 
-{1\over 4} (\cD^+)^2 \, (\x^- (s))^2 
=-{1\over 4} (\cD^-)^2 \, (\x^+ (s))^2 =1~.
\label{fun2}
\ee
One also obtains
\be 
 (\x^+ (s))^2\Big|_{\x=0} = {1\over 4}
\cD^+ \cW \, 
\frac { \sinh^2(s\cM/2)}{(\cM/2)^2} \,
\cD^+ \cW~. 
\label{fun3}
\ee

We have seen that in the case of 
the $U(1)$ background vector multiplet
(\ref{actual-background}), 
any propagator is a superposition
of charged $U(1)$ Green functions, eq. 
(\ref{I+II-ad12}). The same is, of course,  
true for the corresponding heat kernel. 
A heat kernel  of $U(1)$ 
charge $e$, $\bK_{\{e\}}(z,z'|s)$,
 is obtained from 
(\ref{heatkernel1}) by replacing 
$\cW \to e \,W$, ${\bar \cW} \to e\,{\bar W}$,
and so on.

To compute quantum corrections to the 
holomorphic sector (\ref{holomor})
of the low-energy effective action, 
it is sufficient to use a covariantly constant 
vector multiplet under the additional 
{\it complex} constraint
\be 
{\bar \cD}^i_\ad {\bar \cW}=0~,  \qquad 
\cD^i_\a \cW \neq 0~,  \qquad 
\cM_\a{}^\b = -{1\over 4} \cD^i_{\a} \, \cD^\b_i \,
\cW \neq 0~,
\label{quasi}
\ee
which is similar to the condition of relaxed 
super self-duality \cite{KM6}. For such a vector multiplet,
the gauge field is anti-self-dual.  
Introducing the (anti) self-dual components of 
the field strength $\cF_{ab}$, 
\be 
\cF_\pm = \hf (\cF \mp {\rm i}\, \tilde{\cF})~,
\qquad \widetilde{\cF_\pm } = \pm \,{\rm i} \, \cF_\pm ~,
\ee
with $\tilde{\cF}$ the Hodge-dual of $\cF$, 
we have 
\be
\cF = \cF_- ~ . 
\ee
The heat kernel becomes
\bea
K(z,z'|s) &=& -\frac{\rm i}{(4 \pi s)^2} \, 
{\bf tr} \left( \hf 
\frac{(s\cM/2)^2} 
{ \sinh^2(s\cM/2)} \, \right)
\, {\rm e}^{ \frac{{\rm i}}{4} \r
 \cF \coth (s \cF) \r }
\, \d^4(\x (s) )\, \d^4(\bar{\x}) \, \non \\
&& \times
 {\rm e}^{-{\rm i} s {\bar \cW}\cW} I(z,z')~,
\label{heatkernel2}
\eea
where the trace is computed with respect to 
the spinor  indices. It should be noted that 
the variables $\r^a$ and ${\bar \x}^i_\ad$
appear in (\ref{heatkernel2}) independent of $s$.
An important  feature of the background 
vector multiplet chosen is 
\be
(\cD^+)^4 K(z,z'|s) \Big|_{z'=z} =0~. 
\label{impo}
\ee

If the background vector multiplet is such that 
\be
\cD_A \cD_B \cW =
\cD_A \cD_B {\bar \cW} =0~,
\ee
then the heat kernel drastically simplifies:
 \bea
K(z,z'|s) &=& -\frac{\rm i}{(4 \pi s)^2} \,
\, {\rm e}^{ \frac{{\rm i}}{4s} \r^2}
\, \d^4(\x (s) )\, \d^4( \bar{\x}(s)) \,
 {\rm e}^{-{\rm i} s {\bar \cW}\cW} I(z,z')~, 
\label{heatkernel3}
\eea
where 
\be
\x^a_i (s) = \x^\a_i -{s \over 2} \cD^\a_i \cW~, 
\qquad 
{\bar \x}_\ad^i (s) = {\bar \x}_\ad^i 
+{s \over 2} {\bar \cD}_\ad^i {\bar \cW}~.
\ee
This simple heat kernel turns out to be  
most convenient
for computing $F^4$ quantum corrections
(i.e. the $c$-term in (\ref{leea})).

${}$Finally, a covariantly constant 
background vector multiplet  
\be
\cD_A  \cW =
\cD_A  {\bar \cW} =0
\ee
allows us to describe massive matter multiplets, 
with ${\bar \cW}\cW$ being the mass matrix. 
The corresponding heat kernel generates 
free massive propagators \cite{BBIKO}.

\sect{Evaluation of two-loop supergraphs}

We proceed to computing the two-loop 
quantum corrections to the low-energy 
effective action. There are two types of 
two-loop supergraphs, `fish' diagrams 
and `eight' diagrams, and they are generated by 
cubic and quartic interactions, respectively, 
which arise in  the actions 
(\ref{hyper}), (\ref{gluon}) and (\ref{ghost}).

\begin{figure}[!htb]
\begin{center}
\includegraphics{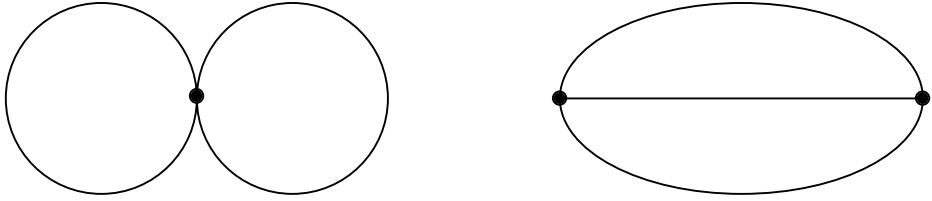}
\caption{Two-loop supergraphs: `eight' diagram and `fish' diagram.}
\end{center}
\end{figure}

\subsection{Hypermultiplet and ghost supergraphs}
We first demonstrate that the two-loop hypermultiplet 
and ghost quantum corrections  cancel each other, 
for an arbitrary on-shell background $\cN=2$
vector multiplet.

The interactions contained in 
(\ref{hyper}) and (\ref{ghost}) are: 
\bea 
S^{(3)}_{{\rm hyper}} &=& 
\int {\rm d}\zeta^{(-4)} \,  
f^{\m \n \l} \,
\breve{q}{}^+_\m \,v^{++}_\n \,
q^+_\l ~, \qquad 
S^{(3)}_{\rm ghost} =
 \int {\rm d}\zeta^{(-4)}\,
 f^{\m \n \l} \,
(\cD^{++} b_\m) \,
v^{++}_\n \,c_\l~,
\eea
with $f^{\m \n \l}$ the structure constants. 
Their combined two-loop contribution, 
\bea
{ {\rm i} \over 2}  \, \Big\langle S^{(3)}_{\rm hyper} \, 
 S^{(3)}_{\rm hyper} \Big\rangle_{\rm 1PI}  
+ { {\rm i} \over 2}  \, \Big\langle S^{(3)}_{\rm ghost} \, 
 S^{(3)}_{\rm ghost} \Big\rangle_{\rm 1PI}  
\equiv \G_{\rm I} ~,
\eea 
can be written down in a rather compact form 
if one notices, following \cite{GIOS1,GIOS2},
\be
\cD^{++} G^{(0,0)} (Z, \,  Z')
= - (u^+u'^-)\,G^{(1,1)} (Z,Z')~. 
\ee
Then one gets
\bea
 \G_{\rm I} &=& 
-\hf \int {\rm d}\zeta^{(-4)}   \int {\rm d}\zeta'^{(-4)} \, 
f^{\m \n \l} \,f^{\m' \n' \l'} \,
\Big\{ 1 + (u^+u'^-)(u^-u'^+) \Big\} \non \\
& & \qquad \qquad
\times  G^{(1,1)}_{\m \, \m'} (Z , Z') \,
G^{(1,1)}_{\n \, \n'} (Z , Z') \, 
G^{(2,2)}_{\l \, \l'} (Z , Z')~. 
\eea 
Use of the identity \cite{GIOS1} 
$1 + (u^+u'^-)(u^-u'^+) = (u^+u'^+)(u^-u'^-)$ gives
\bea
 \G_{\rm I} &=& 
-\hf \int {\rm d}\zeta^{(-4)}   \int {\rm d}\zeta'^{(-4)} \, 
f^{\m \n \l} \,f^{\m' \n' \l'} \,(u^+u'^+)(u^-u'^-)  \non \\
& & \qquad \qquad
\times  G^{(1,1)}_{\m \, \m'} (Z , Z') \,
G^{(1,1)}_{\n \, \n'} (Z , Z') \, 
G^{(2,2)}_{\l \, \l'} (Z , Z')~. 
\label{ghost-hyper}
\eea 
This expression turns out to vanish, 
due to the presence of $(u^-u'^-)$ 
in the integrand, as we are going to show. 

In the case  of an  arbitrary 
on-shell $\cN=2$ vector multiplet, 
the following identity \cite{K} holds
\be 
[\D, (\cD^+)^4] = 0~.
\ee
This idenity implies 
\be
\cD_1^{(ij} \, {\bar \cD}_1^{kl)} \, K(z_1, z_2|s) 
= \cD_2^{(ij} \, {\bar \cD}_2^{kl)} \, K(z_1, z_2|s) ~,
\ee
and the latter in turn leads   \cite{K}
(see also \cite{KM1,KM2}) to 
\bea
(\cD^+)^4 (\cD'^+)^4\, K(z,z'|s) &=& (\cD^+)^4
\Big\{ (u^+u'^+)^4 \,(\cD^-)^4\, 
-{{\rm i}\over 2} (u^+u'^+)^3(u^-u'^+)\, \O^{--} \non \\
&& \qquad \qquad  + (u^+u'^+)^2(u^-u'^+)^2\,\D \Big\} K(z,z'|s)~,
\label{crucial}
\eea
where 
\be
\O^{--} = \cD^{\a\ad} \cD^-_\a \cD^-_\ad 
+\hf \cW (\cD^-)^2 +\hf {\bar \cW}({\bar \cD}^-)^2
+(\cD^-\cW)\cD^- + ({\bar \cD}^-{\bar \cW}){\bar \cD}^-~.
\label{Omega}
\ee
We thus obtain for the $q$-hypermultiplet propagator
\bea
G^{(1,1)} (Z,Z')
&=&  -{\rm i}  (u^+u'^+)
\int_{0}^{\infty} {\rm d}s \, 
(\cD^+)^4 (\cD^-)^4\, K(z,z'|s)   
\non \\
&&  -\hf   (u^-u'^+)
\int_{0}^{\infty} {\rm d}s \, 
(\cD^+)^4 \O^{--} \, K(z,z'|s)  \non \\
&& + {  (u^-u'^+)^2 \over (u^+u'^+)}
(\cD^+)^4 \, \Big\{ {\bf 1} \d^{12}(z-z')\Big\}~. 
\eea
As is seen, only the third term in $G^{(1,1)}(Z,Z')$
is singular, 
in the coincidence limit,  with respect to 
the harmonic variables.
As concerns  the product $(u^+u'^+) \,G^{(1,1)}(Z,Z')$,
it is free of harmonic singularities. Therefore, the only 
harmonic-singular term 
in  the expression 
\be
(u^+u'^+) \, G^{(1,1)}_{\m \, \m'} (Z , Z') \,
G^{(1,1)}_{\n \, \n'} (Z , Z') 
\label{nonsingular}
\ee
is proportional to 
$$
\left(
(\cD^+)^4 \, \Big\{ {\bf 1} \d^{12}(z-z') \Big\} \right)^2~,
$$
and this vanishes on considerations of the Grassmann
algebra. The non-singular part of (\ref{nonsingular}) 
enters $ \G_{\rm I} $  multiplied by 
\be
(u^-u'^-)  \, \d^{(2,-2)} (u,u') =0~.
\ee
Therefore,  $ \G_{\rm I} $ vanishes.

\subsection{Gluon `fish'  supergraph}
The cubic interaction contained in (\ref{gluon}) is
\bea
S^{(3)}_{{\rm SYM}} &=&
{{\rm i} \over 3} \,
{\rm tr}\,\int
{\rm d}^{12}z
\int 
\prod_{a=1}^{3}{\rm d} u_a  \,
\frac{v^{++}(u_1)v^{++}(u_2) v^{++}(u_3)}
{(u^+_1u^+_2)(u^+_2u^+_3)(u^+_3 u^+_1)} \non \\
&=& -{1 \over 6} 
\int {\rm d}^{12}z
\int \prod_{a=1}^{3}{\rm d} u_a  \,
\frac{ f^{\m\n\l} \,
v^{++}_\m(u_1)v^{++}_\n(u_2) v^{++}_\l(u_3)}
{(u^+_1u^+_2)(u^+_2u^+_3)(u^+_3 u^+_1)} ~.
\eea
It generates a two-loop quantum correction, 
\bea
{ {\rm i} \over 2}  \, \Big\langle S^{(3)}_{\rm SYM} \, 
 S^{(3)}_{\rm SYM} \Big\rangle_{\rm 1PI}  
\equiv \G_{\rm II} ~,
\eea 
which can be represented in the form: 
\bea
 \G_{\rm II} &=& -{1\over 12} 
\int {\rm d}^{12}z \,{\rm d}^{12}z'
\int \prod_{a=1}^{3}{\rm d} u_a  \,
\prod_{b=1}^{3}{\rm d} u'_b  \,
 f^{\m\n\l} \, f^{\m'\n'\l'}
\non \\
&&\times  
\frac{ 
G^{(2,2)}_{\m \, \m'} (z,u_1 ; z', u'_1) \,
G^{(2,2)}_{\n \, \n'} (z, u_2; z', u'_2) \, 
G^{(2,2)}_{\l \, \l'} (z,u_3 ; z', u'_3)}
{(u^+_1u^+_2)(u^+_2u^+_3)(u^+_3 u^+_1)
\,(u'^+_1u'^+_2)(u'^+_2u'^+_3)(u'^+_3 u'^+_1)} ~.
\eea
${}$For the $U(1)$ gauge background chosen, 
eq. (\ref{actual-background}),
this can be shown, using the group-theoretical 
results of \cite{KM5,KM7}, to reduce to 
\bea
 \G_{\rm II} &=& -{1\over 2} N(N-1)  
\int {\rm d}^{12}z \,{\rm d}^{12}z'
\int \prod_{a=1}^{3}{\rm d} u_a  \,
\prod_{b=1}^{3}{\rm d} u'_b  \, 
\non \\
&&\times  
\frac{  \bG^{(2,2)}_{\{0\}} (z,u_1 ; z', u'_1) \,
\bG^{(2,2)}_{\{e\}} (z, u_2; z', u'_2)  \, 
\bG^{(2,2)}_{\{-e\}} (z,u_3 ; z', u'_3)}
{(u^+_1u^+_2)(u^+_2u^+_3)(u^+_3 u^+_1)
\,(u'^+_1u'^+_2)(u'^+_2u'^+_3)(u'^+_3 u'^+_1)} ~.
\label{gluonfish}
\eea

The expression (\ref{gluonfish}) 
turns out  to be free of coinciding harmonic 
sigularities, therefore it is safe to use the short form 
of the gluon propagator,
\bea
\bG^{(2,2)}_{\{e\}} (z, u; z', u')  
&=&  
(\cD^+)^4 \bG_{\{e\}} (z,z')  
\, \d^{(-2,2)} (u,u') 
~,\non \\
\bG_{\{e\}} (z,z') &=&
{\rm i}  
\int_{0}^{\infty} {\rm d}s \,
\bK_{\{e\}} (z,z'|s) ~.
\eea
This allows us to rewrite 
$ \G_{\rm II} $ as follows:
\bea
 \G_{\rm II} &=& -{1\over 2} N(N-1)  
\int {\rm d}^{12}z \,{\rm d}^{12}z'
\int \prod_{a=1}^{3}{\rm d} u_a  \,
 \non \\
& &\times 
\frac{  \Big\{ (\cD_1^+)^4 \bG_{\{0\}} (z,z')  \Big\}
 \Big\{ (\cD_2^+)^4 \bG_{\{e\}} (z,z')  \Big\}
(\cD_3^+)^4 \bG_{\{-e\}} (z,z') }
{(u^+_1u^+_2)^2(u^+_2u^+_3)^2(u^+_3 u^+_1)^2 }
\non \\
&=& -{1\over 2} N(N-1)  
\int {\rm d}^{12}z \,{\rm d}^{12}z'
\int \prod_{a=1}^{3}{\rm d} u_a  \,
 \bG_{\{0\}} (z,z') 
 \non \\
& &\times 
(\cD_1^+)^4 \,
\frac{  
 \Big\{ (\cD_2^+)^4 \bG_{\{e\}} (z,z')  \Big\}
(\cD_3^+)^4 \bG_{\{-e\}} (z,z') }
{(u^+_1u^+_2)^2(u^+_2u^+_3)^2(u^+_3 u^+_1)^2 }
~,
\label{gluonfish2}
\eea
where the operators $(\cD^+_a)^4$,  with $a=1,2,3$,
correspond to the same 
superspace point $z$, but different 
harmonics $u^+_a$. 
The propagator $ \bG_{\{0\}} (z,z') $
in this expression is free massless,
\be
 \bG_{\{0\}} (z,z') = 
\int_{0}^{\infty} 
\frac{{\rm d}s}{(4 \pi s)^2} \,
\, {\rm e}^{ \frac{{\rm i}}{4s} \r^2}
\, \d^8(\q-\q')~, \qquad 
\d^8(\q-\q') =
\d^4(\x )\, \d^4( \bar{\x}) ~.
\ee
Using the
Grassmann delta-function contained in 
 $ \bG_{\{0\}} (z,z') $, we can do one of the 
two Grassmann integrals in (\ref{gluonfish2}), say, 
the integral over $\q'$. 
After that, the integrand obtained 
involves 
the following expression
\be
(\cD_1^+)^4 \left(
 \Big\{ (\cD_2^+)^4 \bG_{\{e\}} (z,z')  \Big\}
(\cD_3^+)^4 \bG_{\{-e\} (z,z') }\right)
\Big|_{\x={\bar \x}=0}~.
\label{coinc}
\ee

We are interested in computing holomorphic 
quantum corrections of the form (\ref{holomor}).
It is therefore sufficient to use 
the heat kernel (\ref{heatkernel2}) which involves 
a Grassmann delta-function $\d^4( \bar{\x}) $. 
Since each of the two propagators in (\ref{coinc}) 
contains such a delta-function,
and since 
\be
{\bar D}^n\, \d^4 ({\bar \x})|_{{\bar \x}=0}  = 0~,
\quad n<4~, \qquad \quad
{\bar D}^4
\, \d^4 ({\bar \x}) =1 ~,
\label{barr}
\ee
one  needs at least eight derivatives 
$\bar \cD$'s in order to annihilate 
the delta-functions in the coincidence limit.
However $(\cD^+)^4 = {1 \over 16} (\cD^+)^2 \, 
 ({\bar \cD}^+)^2$,  and therefore
(\ref{coinc}) contains only six derivatives
$\bar \cD$'s. We thus conclude that 
$\G_{\rm II}$ does not produce quantum corrections 
of the form (\ref{holomor}).
This clearly disagrees with \cite{BPT}.

\subsection{Gluon `eight'  supergraph}
The quartic interaction contained in (\ref{gluon}) is
\bea
S^{(4)}_{{\rm SYM}} &=&
{1 \over 4} \,
{\rm tr}\,\int
{\rm d}^{12}z
\int \prod_{a=1}^{4}
{\rm d} u_a  
\frac{v^{++}(u_1)v^{++}(u_2) v^{++}(u_3) v^{++}(u_4)}
{(u^+_1u^+_2)(u^+_2u^+_3)(u^+_3 u^+_4)
(u^+_4 u^+_1)}~.
\eea
It generates
the two-loop quantum correction 
\bea
&& \G_{\rm III} =  
 \Big\langle S^{(4)}_{\rm SYM} \,  \Big\rangle
= - \hf \, {\rm tr} (T^\m T^\n T^\l T^\s)
\int {\rm d}^{12}z
\int \prod_{a=1}^{4} u_a 
\label{gluoneight}\\
 && \times 
\left\{
\frac{G^{(2,2)}_{\m \, \n} (z,u_1 ; z, u_2) \,
G^{(2,2)}_{\l \, \s} (z, u_3; z, u_4) }
{(u^+_1u^+_2)(u^+_2u^+_3)(u^+_3 u^+_4)
(u^+_4 u^+_1)} 
+ \hf 
\frac{ G^{(2,2)}_{\m \, \l} (z,u_1 ; z, u_3) \,
G^{(2,2)}_{\n \, \s} (z, u_2; z, u_4) }
{(u^+_1u^+_2)(u^+_2u^+_3)(u^+_3 u^+_4)
(u^+_4 u^+_1)} \right\} ~.
\non
\eea 

The second term in 
(\ref{gluoneight}) 
can be seen to be free of coinciding harmonic 
sigularities. 
When  computing this term, it is 
 therefore  safe to use the short form 
of $G^{(2,2)}$, eq. (\ref{short}), for the two Green functions
involved.
We are only interested in computing holomorphic 
quantum corrections of the form (\ref{holomor}).
It is therefore sufficient to use 
the heat kernel (\ref{heatkernel2}) which involves 
a Grassmann delta-function $\d^4( \bar{\x}) $.
Due to (\ref{barr}), it is then clear that   
the second term in  (\ref{gluoneight}) 
does not generate any quantum corrections
of the form (\ref{holomor}).

The remaining part of the  supergraph  
under consideration
\bea
 \G'_{\rm III} &=&  
- \hf \, {\rm tr} (T^\m T^\n T^\l T^\s)
\int {\rm d}^{12}z
\int \prod_{a=1}^{4} u_a
\frac{G^{(2,2)}_{\m \, \n} (z,u_1 ; z, u_2) \,
G^{(2,2)}_{\l \, \s} (z, u_3; z, u_4) }
{(u^+_1u^+_2)(u^+_2u^+_3)(u^+_3 u^+_4)
(u^+_4 u^+_1)} 
\label{gluoneight2}
\eea
contains coinciding harmonic 
sigularities, and therefore we have to use 
the longer form of $G^{(2,2)}$, eq. (\ref{longer}), 
in order to resolve the singularities.

In accordance with eq. (\ref{crucial}), we  have
\bea
&&  \frac{(\cD^+)^4 (\cD'^+)^4\, K(z,z'|s)} {(u^+u'^+)} \,
(\cD^{--})^2 \, \d^{(2,-2)} (u,u') \non \\
&& \qquad 
= (\cD^+)^4(\cD^-)^4\, K(z,z'|s) \, 
 (u^+u'^+)^3  (\cD^{--})^2 \, \d^{(2,-2)} (u,u') \non \\
&& \qquad \quad
-{{\rm i}\over 2} (\cD^+)^4\O^{--} \, K(z,z'|s)
(u^-u'^+) (u^+u'^+)^2(\cD^{--})^2 \, \d^{(2,-2)} (u,u') 
\non \\
&&  \qquad \quad
+  (\cD^+)^4 \D  K(z,z'|s)\,
(u^-u'^+)^2 (u^+u'^+) (\cD^{--})^2 \, \d^{(2,-2)} (u,u') 
\label{crucial2}
\eea
Since $\cD^{--} (u^+u'^+)= (u^-u'^+)$ and
$(u^+u'^+) \, \d^{(2,-2)} = 0$, 
the first term on the right 
vanishes. 
The rest can be rewritten as follows 
\bea
&&  \frac{(\cD^+)^4 (\cD'^+)^4\, K(z,z'|s)} {(u^+u'^+)} \,
(\cD^{--})^2 \, \d^{(2,-2)} (u,u') \non \\
&& \qquad =- {\rm i}  (\cD^+)^4
\left\{ 
\O^{--} \, K(z,z'|s)
-2 {\pa \over \pa s}     
K(z,z'|s) \, \cD^{--} \right\}  \d^{(-1,1)}(u,u')~, 
\label{crucial3}
\eea
where we have used the heat kernel equation 
and the identity 
$(u^-u'^+)^3 \, \d^{(2,-2)} (u,u') = \d^{(-1,1)} (u,u')$. 
We are only interested in computing holomorphic 
quantum corrections of the form (\ref{holomor}).
It is therefore sufficient to use 
the heat kernel (\ref{heatkernel2}). 
Since the relation (\ref{impo}) holds
for this heat kernel, 
from (\ref{crucial3})  we deduce
\bea
&&  \frac{(\cD^+)^4 (\cD'^+)^4\, K(z,z'|s)} {(u^+u'^+)} \,
(\cD^{--})^2 \, \d^{(2,-2)} (u,u') \Big|_{z'=z} \non \\
&& \qquad \qquad \qquad
=- {\rm i} \, \d^{(-1,1)}(u,u') \,
 (\cD^+)^4\O^{--} \, K(z,z'|s)\Big|_{z'=z}~.
\label{crucial4}
\eea
This leads to 
\bea
 \G'_{\rm III} &=&  
 \hf \, {\rm tr} (T^\m T^\n T^\l T^\s)
\int {\rm d}^{12}z
\int {\rm d} u_1  {\rm d} u_2 \,
\frac{\cL^{++}_{\m\n}(z,u_1) \, \cL^{++}_{\l\s}(z,u_2)}
{(u^+_1u^+_2)^2}~, 
\label{gluoneight3}
\eea
where 
\be
\cL^{++}(z,u) = { {\rm i} \over 2}
 \int_{0}^{\infty} {\rm d}s \,  s \,
(\cD^+)^4\O^{--} \, K(z,z'|s)\Big|_{z'=z}~.
\label{L++}
\ee
${}$Finally, specifying (\ref{gluoneight3}) 
to the case of the $U(1)$ background 
(\ref{actual-background}) gives
\bea
 \G'_{\rm III} &=&  
 \hf \, N(N-1)
\int {\rm d}^{12}z
\int {\rm d} u_1  {\rm d} u_2\,
\frac{\bL^{++}_{\{e\}}(z,u_1) \, 
\bL^{++}_{\{e\}}(z,u_2)}
{(u^+_1u^+_2)^2}~. 
\label{gluoneight4}
\eea
Here $\bL^{++}_{\{e\}}(z,u)$ is obtained 
from $\cL^{++}(z,u) $ by  replacing 
$K(z,z'|s) \to \bK_{\{e\}}(z,z'|s)$.

It is almost trivial to calculate 
$\cL^{++}(z,u)$ for 
the heat  kernel (\ref{heatkernel2}).
Indeed, since 
$$ 
(\cD^+)^4 = {1 \over 16} (\cD^+)^2 \, 
 ({\bar \cD}^+)^2 ~, 
$$
and since the heat  kernel (\ref{heatkernel2})
involves the  Grassmann delta-function 
$$
\d^4 ({\bar \x}) =  ({\bar \x}^+)^2 \, ({\bar \x}^-)^2~, 
$$ 
with the properties 
$${\bar \cD}^n\, \d^4 ({\bar \x})|_{{\bar \x}=0}  = 0~,
\quad n<4~, \qquad \quad
{1 \over 16} ({\bar \cD}^+)^2 \, 
 ({\bar \cD}^-)^2 
\, \d^4 ({\bar \x}) =1 ~,
$$
it is only the third term in the operator $\O^{--}$, 
eq. (\ref{Omega}), which provides a non-vanishing 
contribution to $\cL^{++}(z,u)$.  
It then  remains to evaluate
$$
{1\over 4} (\cD^+)^2 \,
\Big\{\d^4(\x(s)) \,I(z,z')\Big\} \Big|_{z'=z}~, 
\qquad \d^4(\x(s))=(\x^+(s))^2 \,(\x^-(s))^2~.
$$
Using the  properties of the parallel 
displacement propagator 
(see Appendix C)
\be
I(z,z) ={\bf 1}~, \qquad 
\cD^+_\a I(z,z')\Big|_{z'=z} 
= (\cD^+)^2 I(z,z') \Big|_{z'=z} =0~, 
\ee
and the identities (\ref{fun1})--(\ref{fun3}), 
we obtain 
\be
\cL^{++}(z,u) = {1 \over 4(4\p)^2} (\cD^+ \cW)^2 \, 
{\bar \cW}  \int_{0}^{\infty} {\rm d}s \,  s \,
 {\rm e}^{- s {\bar \cW}\cW}~,
\label{LL}
\ee
where 
the proper-time integral has been Wick-rotated.
The expression for $\cL^{++}$ obtained 
is so simple due to the relation 
\be
{\bf det} \left( \frac{ s \cF}
{\sinh  ( s \cF) }
\right)^{1 \over 2} \,(\x^+(s))^2
\Big|_{\x =0} = {s^2 \over 4}  \, (\cD^+ \cW)^2 
\ee
that holds 
for the heat kernel (\ref{heatkernel2}).

Replacing $\cW \to e \,W$ in (\ref{LL}) gives
\be 
\bL^{++}_{\{e\}}(z,u)
={1 \over 4(4\p)^2 } \,\sqrt{{N-1 \over N}  } \,
\frac{ (D^+ W)^2 } {W^2 
{\bar W}  } ~,
\label{LLfinal}
\ee
where we have inserted the expression for $e$.
Now, eqs. (\ref{holomor}), (\ref{gluoneight4})
and (\ref{LLfinal})
lead to the final result (\ref{l-two})

\sect{Discussion}
Using the $\cN=2$ off-shell formulation for 
$\cN=4$ SYM in harmonic superspace, 
in this paper we computed the two-loop 
quantum correction to the holomorphic sector
$\L ({\bar \J}) $ of the low-energy effective 
action (\ref{leea}). Relation (\ref{l-two}) is
our main result. 
As explained in section 1, 
the explicit structure of  $\L_{\rm two-loop}({\bar \J}) $
is consistent with the self-duality equations
(\ref{sde1}) and (\ref{sde2}), as well as 
with the conjectured 
correspondence between the D3-brane action 
in $AdS_5 \times S^5$ and 
the low-energy effective action for $\cN=4$ $SU(N)$ 
SYM with the gauge group  spontaneously 
broken to $SU(N-1) \times U(1)$. 

It is a by-product of our calculation that no quantum 
correction to the $c$-term in (\ref{leea}) is 
generated in $\cN=4$ SYM at the two-loop level.
This is consistent with the Dine-Seiberg 
non-renormalization conjecture \cite{DS} and 
its generalized form  \cite{KM5,KM7}.
The absence of two-loop $F^4$ quantum corrections 
was  also established in \cite{KM7} using the 
$\cN=1$ superfield formulation for $\cN=4$ SYM.
It should be pointed out that two-loop $F^4$ 
contributions 
are actually generated from both the hypermultiplet 
and ghost `fish' supergraphs. 
But in the case of $\cN=4$ SYM,
the  hypermultiplet and ghost  contributions 
cancel each other at two loops,
in accordance with the consideration in subsection 4.1.

Apart from the analysis given in the present paper, 
two independent  calculations 
of the two-loop $F^6$ quantum correction 
in  $\cN=4$ SYM have been carried out
\cite{BPT,KM7}. They  were based, respectively, 
on the use of (i) $\cN=2$  harmonic supergraphs 
\cite{BPT};  and (ii) $\cN=1$  supergraphs  
\cite{KM7}. Unfortunately, 
there is no agreement between the results
of the three calculations given.

It is not difficult 
to compare the outcome
of our calculation with that of \cite{BPT}, 
as well as to  figure out the origin 
of differences. Referring to eq. 
(\ref{kappa}),  the result of our calculation corresponds 
to the choice $\k_1$, while 
the calculation in \cite{BPT}  leads to $\k_2$. 
Comparing the technical blocks of our calculation with 
\cite{BPT}, the results  prove to differ in the sectors 
of the `fish' and `eight'  gluon diagrams. In our approach, 
there is no contribution from the `fish,' and the entire 
quantum correction comes from the `eight' diagram.
According to \cite{BPT}, the situation is just opposite --
the entire quantum correction 
comes from the `fish' supergraph.
We believe that the calculation in \cite{BPT} is 
not  correct.
When evaluating two-loop harmonic supergraphs, 
it turns out that 
a crucial role is played by the identity 
(\ref{crucial}) (or its equivalent form 
originally given in \cite{KM1,KM2})
which allows one to resolve coinciding harmonic 
singularities. Although this identity 
had been known at the time, it was not taken 
into account in \cite{BPT}. Unfortunately, the 
notoriously difficult 
{\it problem of coinciding harmonic singularities} 
\cite{GKS,BK,KM1,KM2,KM8} is the high price 
one has to pay for the indispensable advantage 
of  manifest $\cN=2$ supersymmetry 
in the quantum harmonic superspace approach. 
This is the technical problem which has no analogue
in $\cN=1$ superspace and which all 
quantum practitioners face, sooner or later,
once engaged  with harmonic supergraphs   
(the founding fathers of harmonic superspace 
have only considered  a few simple examples 
of singular harmonic supergraphs
\cite{GKS,GIOS2}).
At the one-loop level, the problem 
of coinciding harmonic singularities was solved 
in  \cite{BK,KM1,KM2}.  But it then emerged again at two 
loops \cite{BPT,KM8}.  Now we believe, on the basis
of the experience gained and 
the analysis  given in this paper and also in \cite{KM8},  
that the two-loop approximation is under control. 
However,  three loops may bring 
a new challenge. 
It would, therefore,  be very important 
to formulate well-defined rules 
to resolve coinciding singularities 
in  harmonic supergraphs at any loop order.

According to the $\cN=1$ supergraph 
considerations in \cite{KM7}, no two-loop $F^6$ 
quantum correction is generated in $\cN=4$ SYM. 
This is not compatible with  the correspondence between 
the D3-brane action in $AdS_5 \times S^5$
and the low-energy action for $\cN=4$ SYM, 
provided one naively identifies the component 
fields of the $\cN=2$ vector multiplet 
$(\f, W_\a)$ with those living on the D3-brane.
But most likely, such an identification is not consistent,  
as was discussed in the Introduction.

The two-loop $F^6$ results of \cite{KM7} and 
of the present paper do not match so far. 
This discrepancy may be due to the fact that, 
in the process of background-field quantization,  
different gauge conditions  were used
in  \cite{KM7} and in the present paper
(in particular, the gauge condition in \cite{KM7}
respected only the  $\cN=1$ supersymmetry).
Unlike the $S$-matrix, which is determined by 
the effective action at its stationary point,  
the off-shell effective action in gauge theories  
is known to depend on the choice 
of gauge conditions. 
As shown, e.g.,  in \cite{FK}, the one-loop effective action 
evaluated at a classical stationary point
(the on-shell $U(1)$ vector multiplet in our case) 
 is gauge independent. At the same classical stationary 
point, the two-loop effective action becomes 
gauge independent only after adding up some
one-particle reducible functional generated by 
the one-loop action. For different gauge conditions, 
the corresponding effective actions are related  by 
a field redefinition. The $F^6$ term 
and higher order terms may be sensitive 
to such field redefinitions. Due to its importance, 
let us try to make this point a bit more precise.

${}$For a gauge theory of Yang-Mills type, 
let $\G[\vf]$ be the gauge-invariant effective action 
defined in the framework of background-field
quantization, 
\be
\G[\vf] = S[\vf] +\sum_{L=1}^{\infty}
\hbar^L \G^{(L)}[\vf]~, 
\ee
with $S[\vf]$ the classical action, 
and $\G^{(L)}[\vf]$ the $L$-loop quantum correction.
At a generic point $\vf^i$, the effective action 
explicitly depends on the gauge conditions 
used in the process of 
the Faddeev-Popov quantization. 
This dependence disappears at the stationary 
point of the effective action 
\be
\G_{, i} [\hat{\vf}] =0~, \qquad 
\hat{\vf }= \vf_0 + \hbar \vf^{(1)} 
+ \hbar^2 \vf^{(2)}  + \cdots~, 
\ee
with $\vf_0$ the classical stationary point, 
$S_{, i} [\vf_0] =0$. 
More precisely, it is $\G [\hat{\vf}] $ which 
does not depend on the gauge conditions 
chosen, but both $\G[\vf]$ and $\hat{\vf}$ 
are ``gauge-dependent.''
It can be shown \cite{FK} that 
the one-loop effective action evaluated 
at the classical stationary point, 
$ \G^{(1)}[\vf_0]$,  does not depend 
on the gauge conditions chosen, but this is 
no longer true for the two-loop quantum correction
$ \G^{(2)}[\vf_0]$. 
Now, in the case of $\cN=4$ SYM, 
the  corresponding 
formulations in $\cN=1$ superspace 
and in $\cN=2$ superspace require 
different gauge conditions. For this theory,
there exists  perfect agreement between 
$ \G^{(1)}[\vf_0]$
evaluated in terms of $\cN=1$ superfields
\cite{BKT} and $\cN=2$ superfields \cite{KM1}.
But the two-loop quantum corrections 
should not necessarily coincide.
Their relationship will be studied in 
a separate publication.

Of course, if there exists an exact solution to 
all orders, $\hat{\vf} = \vf_0$, then $\G [\vf_0] $ 
does not depend on the gauge conditions 
chosen.  This is what happens in QED where 
any constant field configuration, 
$F_{mn}={\rm const}$, is 
an exact solution for the Euler-Heisenberg 
effective action.
The situation is different 
in the case  of the $\cN=4$ SYM  
effective action (\ref{leea}). 
An $\cN=2$  supersymmetric generalization of 
$F_{mn}={\rm const}$ is 
\be 
{\bar D}^i_\ad W = D^i_\a {\bar W} 
=D^{ij} W = {\bar D}^{ij} {\bar W} 
= \pa_a W = \pa_a {\bar W} =0~,
\label{tree}
\ee
with  $F$ associated with 
the component
$D^i{}_\a D_{\b i} W|_{\q =0}$
and its conjugate.  
Such a superfield 
is a solution for the classical $U(1)$ vector 
multiplet action (\ref{classical}).
However, once we replace $S_{\rm cl}$ 
by the action  (\ref{leading}), which contains 
the leading quantum correction, 
the configuration (\ref{tree}) is  no longer a solution. 
This can be seen  by inspecting  the equation 
of motion
\be 
(D^+)^2 F'(W) +({\bar D}^+)^2 {\bar F}'({\bar W}) 
+ (D^+)^4 \left\{ 
({\bar D}^-)^2 
\frac{\pa }{\pa W} + (D^-)^2 
\frac{\pa }{\pa \bar W}  \right\} H( W , {\bar W} )= 0
\label{quant-eq-mo}
\ee 
corresponding to the action  
\be
I =  2\,{\rm Re}  
\int {\rm d}^8z \,  F(W)  
+ \int {\rm d}^{12}z \,  H( W , {\bar W})~.
\ee
Thus, quantum corrections deform the 
classical superfield solution (\ref{tree}).
This is actually expected keeping in mind
the correspondence between the D3-brane action 
(\ref{D3}) and  the $\cN=4$ SYM  effective action. 
Indeed, it was mentioned in the Introduction 
that the only constant field solutions 
for the D3-brane in $AdS_5 \times S^5$ 
are $F_{mn}=0$ and $X^I ={\rm const}$, 
although the quadratic part of the D3-brane 
action (\ref{D3}) possesses more general solutions:
$F_{mn}={\rm const} $ and $X^I ={\rm const}$. 

In fact, there exists a simple exact solution for 
the low-energy effective action (\ref{leea}):
\be
{\bar D}^i_\ad W =  D^{ij} W 
= \pa_a W =0~, \qquad {\bar W} = {\rm const}~.
\label{complex-solut}
\ee
Its use should  in principle 
be enough to  uniquely restore
the holomorphic $\L$-term in (\ref{leea})
which clearly indicates that this $\L$-term 
is  gauge-independent. 
It is important to note that such a solution allows us to 
extract nontrivial information about the effective action 
only in the $\cN=2$ superspace setting (when reduced
to $\cN=1$ superspace, the effective action trivializes 
unless we relax the constraints on $\bar W$ to 
${\bar D}^i_\ad {\bar W} 
\neq 0$).  This may be the origin of the discrepancy 
between the $\cN=1$  \cite{KM7} and $\cN=2$ results. 
In Minkowski space, the conditions 
(\ref{complex-solut}) are of course purely formal, 
as they are obviously inconsistent with 
the structure of  a single real vector multiplet. 
But the use of complex field configurations 
is completely legitimate when 
we are interested in special sectors of the
effective action, see, e.g.,   \cite{DI}.

\vskip.5cm

\noindent
{\bf Acknowledgements:}\\
It is the pleasure to thank Ian McArthur for 
numerous discussions and for 
collaboration at early stages of this project.
The author is grateful to Ioseph Buchbinder 
and especially to Arkady Tseytlin for reading the manuscript 
and for important critical comments and suggestions.
This work is supported in part by the Australian Research
Council. 

\begin{appendix}

\sect{Superspace integration measures}
Here we introduce  various $\cN=2$ superspace 
integration measures used throughout this paper. 
They are defined
in terms of the spinor covariant derivatives
$D^i_\a$ and ${\bar D}^\ad_i$, with 
$i =\hat{1},  \hat{2}$, and the related 
fourth-order operators
\bea
 D^4 = {1\over 16} (D^{\hat{1}} )^2 (D^{\hat{2}})^2 
=  {1\over 48} D^{ij}  D_{ij}~, 
\qquad 
{\bar D}^4 = {1\over 16} 
({\bar D}_{\hat{1}} )^2 ({\bar D}_{\hat{2}})^2 
=  {1\over 48} {\bar D}^{ij}  {\bar D}_{ij}~, 
\eea
where the second-order operators 
$D^{ij}$ and ${\bar D}^{ij}$ are given in 
(\ref{ij}).
Integration over the chiral subspace is defined by 
\be 
\int {\rm d}^8 z \, L_{\rm c} = 
\int {\rm d}^4 x\, D^4 L_{\rm c}~, 
\qquad {\bar D}^i_\ad L_{\rm c} =0~.
\ee
Integration over the full superspace 
is defined by 
\be 
\int {\rm d}^{12} z \, L = 
\int {\rm d}^4 x\, {\bar D}^4 D^4 L~. 
\ee
In terms of the harmonic-dependent 
spinor derivatives $D^\pm_\a = D^i_\a \,u^\pm_i$ 
and ${\bar D}^\pm_\ad = {\bar D}^i_\ad \,u^\pm_i$, 
and the related fourth-order operators 
\be 
(D^+)^4 = {1 \over 16} (D^+)^2 ({\bar D}^+)^2 ~, 
\qquad  
(D^-)^4 = {1 \over 16} (D^-)^2 ({\bar D}^-)^2 ~, 
\ee
integration over the analytic subspace is defined by
\be 
\int {\rm d}\z^{(-4)} \, L^{(+4)} = 
\int {\rm d}^4 x \int {\rm d}u\, (D^-)^4 L^{(+4)}~, 
\qquad 
D^+_\a L^{(+4)} =
{\bar D}^+_\ad L^{(+4)} =0~.
\ee
Integration over the group manifold 
$SU(2)$ is defined according to \cite{GIKOS}
\be 
 \int {\rm d}u \, 1 = 1~\qquad 
 \int {\rm d}u \, u^+_{(i_1} \cdots u^+_{i_n}\,
u^-_{j_1} \cdots u^-_{j_m)} =0~, 
\quad n+m >0~.
\ee

\sect{Group-theoretical results}

Here we describe
the $SU(N)$ conventions adopted in the paper.
Lower-case Latin letters from the middle of the alphabet, 
$i,j,\dots$, 
will be used to denote matrix elements in the fundamental, 
with the convention $i =0,1,\dots, N-1 \equiv 0, \un{ i}$.
We choose a Cartan-Weyl basis 
to consist of the elements:
\be 
H^I = \{ H^0, H^{\un{I}}\}~, \quad  \un{I} = 1,\dots, N-2~, 
\qquad \quad E^{ij}~, \quad i\neq j~. 
\label{C-W}
\ee 
The basis elements in the fundamental representation 
are defined similarly to \cite{Georgi}, 
\bea
(E^{ij})^{kl} &=& \d^{ik}\, \d^{jl}~, \non \\
(H^I)^{kl} &=& \frac{1}{\sqrt{(N-I)(N-I-1)} }
\Big\{ (N-I)\, \d^{kI} \, \d^{lI} - 
\sum\limits_{i=I}^{N-1} \d^{ki} \, \d^{li} \Big\} ~,
\label{C-W-2}
\eea
and are characterized by the properties
\be
{\rm tr} (H^I\,H^J) = \d^{IJ}~, 
\qquad 
{\rm tr} (E^{ij}\,E^{kl}) = \d^{il}\,\d^{jk}~, 
\qquad {\rm tr} (H^I \,E^{kl}) =0~.
\ee
A generic element of the Lie algebra $su(N)$ is 
\be
v = v_I \, H^I + v_{ij} \, E^{ij} \equiv v_\m \,T^\m~,  
\qquad i \neq j ~, 
\label{generic}
\ee
${}$For $SU(N)$, the operation of trace 
in the adjoint representation, 
${\rm tr}_{\rm Ad}$, 
is related to that in the fundamental, 
${\rm tr}$, as follows
\be 
{\rm tr}_{\rm Ad} \,v^2 = 2N \, {\rm tr} \, v^2 ~,
\qquad  v \in su(N)~.
\ee

Since the basis 
(\ref{C-W}) is not orthonormal, 
${\rm tr} (T^\m \, T^\n)  =g^{\m \n} \neq \d^{\m \n}$, 
it is
necessary to keep track of the Cartan-Killing metric when 
working with adjoint vectors. For any elements
$u=u_\m T^\m $ and $v=v_\m T^\m $ of the Lie algebra, 
we have $u\cdot v =  {\rm tr} (u\,v) =u_\m \,v^\m$, 
where $v^\m = g^{\m\n} v_\n$
($v^I =v_I$, $v^{ij}= v_{ji}$). 
The structure constants of $SU(N)$ are defined 
in a standard way, 
$[T^\m , T^\n] = {\rm i} \, f^{\m\n}{}_\l T^\l
=  {\rm i} \, f^{\m\n \l} T_\l$.

\sect{Parallel displacement propagator}
In this appendix we describe, 
basically following \cite{KM3}, 
the main properties of the 
parallel displacement propagator
$I(z,z')$ in $\cN=2$ superspace. 
This object is uniquely specified by 
the following requirements:\\
(i) the gauge transformation law
\be
 I (z, z') ~\to ~
{\rm e}^{{\rm i} \t(z)} \,  I (z, z') \,
{\rm e}^{-{\rm i} \t(z')} ~
\label{super-PDO1}
\ee
with respect to the  gauge ($\t$-frame) transformation 
of  the covariant derivatives (\ref{tau});\\
(ii) the equation  
\be
\X^A \cD_A \, I(z,z') 
= \X^A \Big( D_A +{\rm i} \, \cV_A(z) \Big) I(z,z') =0~;
\label{super-PDO2}
\ee
(iii) the boundary condition 
\be 
I(z,z) ={\bf 1}~.
\label{super-PDO3}
\ee
These imply the important relation
\be
I(z,z') \, I(z', z) = {\bf 1}~,
\label{collapse}
\ee
as well as 
\be
\X^A \cD'_A \, I(z,z') 
= \X^A  \Big( D'_A \,I(z,z') 
 - {\rm i} \, I(z,z') \, \cV_A(z') \Big) =0~.
\ee

Let $\J(z)$ be a {\it harmonic-independent}
superfield transforming in some representation of 
the gauge group, 
\be 
\J(z) ~\to ~ {\rm e}^{{\rm i} \t(z)} \, \J(z)~.
\ee
Then it can be represented by 
the  covariant Taylor series \cite{KM3} 
\be 
\J(z) = I(z,z') \,\sum_{n=0}^{\infty} {1 \over n!}\,
\X^{A_n} \ldots \X^{A_1} \, 
\cD'_{A_1} \ldots \cD'_{A_n} \, \J(z') ~. 
\label{super-Taylor2}
\ee

The fundametal properties of the
parallel displacement propagator are 
\cite{KM3}
\bea 
\cD_B I(z,z') &=& {\rm i} \, I(z,z') \,
\sum_{n=1}^{\infty} { 1  \over (n+1)!} \,
\Big\{
n \, \X^{A_n} 
\ldots \X^{A_1}  
\cD'_{A_1} \ldots \cD'_{A_{n-1} } \cF_{A_n \,B } (z') 
\label{super-PTO-der1} \\
&+& \hf  
(n-1)\, 
\X^{A_n} T_{A_n \,B}{}^C \,\X^{A_{n-1}} 
\ldots \X^{A_1}  
\cD'_{A_1} \ldots \cD'_{A_{n-2} } \cF_{A_{n-1} \,C } (z') \Big\}~,
\non 
\eea 
and
\bea 
\cD_B I(z,z') &=& {\rm i} \,
\sum_{n=1}^{\infty} {(-1)^{n} \over (n+1)! } \, \Big\{
- \X^{A_n} 
\ldots \X^{A_1}  
\cD_{A_1} \ldots \cD_{A_{n-1} } \cF_{A_n \,B } (z) 
\label{super-PTO-der-mod} \\
&+& \hf  
(n-1)\,
\X^{A_n} T_{A_n \,B}{}^C \,\X^{A_{n-1}} 
\ldots \z^{A_1}  
\cD_{A_1} \ldots \cD_{A_{n-2} } \cF_{A_{n-1} \,C } (z) \Big\}
\, I(z,z')~.
\non 
\eea
Here $\cF_{AB}$ denotes the superfield strength defined
as follows
\be
[\cD_A , \cD_B \} = T_{AB}{}^C \cD_C 
+{\rm i}\, \cF_{AB}~.
\ee
In the case of a covariantly constant vector multiplet, 
the series in (\ref{super-PTO-der1}) and
(\ref{super-PTO-der-mod}) terminate as only the tensors
$\cF_{CD}$, $\cD_A\cF_{CD}$ and $\cD_A \cD_B \cF_{CD}$
have  non-vanishing components.

The parallel displacement propagator
in the $\l$-frame is obtained from that in 
the $\t$-frame by the transformation 
\be
I(z,z') \equiv I_\t(z,z') 
\quad \longrightarrow \quad 
{\rm e}^{{\rm i}\O(z,u)} \, I(z,z') \,
{\rm e}^{-{\rm i}\O(z',u')}
\equiv I_\l(Z , Z')~,
\ee 
with $Z=\{z,u\}$.

\sect{Free propagators}
${}$Following \cite{Sok},
introduce a generalization of the 
supersymmetric two-point function $\r(z,z')$, 
eq.  (\ref{super-interval}),
\bea
\hat{\r}^a &=&  \hat{\r}^a (Z,Z') 
= - \hat{\r}^a (Z',Z)   
= \r^a - 2{\rm i} \, 
\frac{u^+_{(i} u'^+_{j)} \, 
\x^i \s^a {\bar \x}^j } {(u^+u'^+)}~.
\label{ranal}
\eea 
It follows from  (\ref{trivial}) 
\be 
D^+_\b \hat{\r}^a = D'^+_\b \hat{\r}^a =0~, 
\qquad 
{\bar D}^+_\bd \hat{\r}^a
={\bar D}'^+_\bd \hat{\r}^a=0~.
\ee
Thus the two-point function introduced is analytic 
with respect to $Z$ and $Z'$, and therefore it can 
be represented in the form 
$ \hat{\r}^a = \hat{\r}^a (\z,\z') 
= - \hat{\r}^a (\z',\z) $, where 
the  variables $ \z= (y^a,\q^{+\a},{\bar\q}^+_{\dot\a}, \,
u^+_i,u^-_j) $ parametrize the analytic subspace
of  harmonic superspace.
The result can be found, 
e.g., in \cite{GIOS2}. 

With the use of (\ref{ranal}),
the free massless heat kernel can be written 
in two equivalent form
 \bea
K(z,z'|s) &=& -\frac{\rm i}{(4 \pi s)^2} \,
\, {\rm e}^{ \frac{{\rm i}}{4s} \r^2}
\, \d^4(\x )\, \d^4( \bar{\x})
=  -\frac{\rm i}{(4 \pi s)^2} \,
\, {\rm e}^{ \frac{{\rm i}}{4s} \hat{\r}^2}
\, \d^4(\x )\, \d^4( \bar{\x})~.
\eea
Since $\hat{\r}^2$ is analytic at $Z$ and $Z'$, 
we obtain 
\bea
(D^+)^4 ( D'^+)^4 \,K(z,z'|s)
&=& -\frac{\rm i}{(4 \pi s)^2} \,
\, {\rm e}^{ \frac{{\rm i}}{4s} \hat{\r}^2}
(D^+)^4 (D'^+)^4
\, \d^4(\x )\, \d^4( \bar{\x}) \non \\
&=& 
-\frac{\rm i}{(4 \pi s)^2} \,
\, {\rm e}^{ \frac{{\rm i}}{4s} \hat{\r}^2}
(u^+u'^+)^4~.
\eea
The free massless $q$-hypermultiplet 
propagator becomes 
\bea
G^{(1,1)} (Z,Z')
&=&  -{\rm i}  
\int_{0}^{\infty} {\rm d}s \, 
(D^+)^4 (D'^+)^4\, {K(z,z'|s)   
 \over (u^+ u'^+)^3}
= - {{\rm i} \over 4\p^2 }\,
{ (u^+ u'^+) \over 
\hat{\r}^2 +{\rm i} \ve  }
\Big|_{\ve \to +0} ~,
\label{d5}
\eea 
and the latter expression for $G^{(1,1)} $ was given, 
e.g.,  in \cite{GIOS2}.
A similar, although somewhat uglier, representation 
follows for the gluon propagator. 

Representation (\ref{d5}) is known to be
extremely useful for computing correlation
functions of composite operators in 
the unbroken phase of $\cN=2, 4$ 
superconformal field theories, 
see \cite{GIOS2} and references therein.
It is not clear whether such ideas might be 
of any use in  the case of 
background-field calculations.

\end{appendix}

\end{document}